\documentclass[a4paper,11pt]{article}

\usepackage{jheppub} 

\usepackage[T1]{fontenc} 

\usepackage{amsmath, amssymb, bm, graphicx, graphics, color, mathrsfs}

\newcommand{\red}[1]{\textcolor[rgb]{1.00,0.00,0.00}{#1}}

\newcommand{\be}{\begin{equation}}
\newcommand{\ee}{\end{equation}}
\newcommand{\ben}{\begin{eqnarray}}
\newcommand{\een}{\end{eqnarray}}

\newcommand{\la}{{\lambda}}

\newcommand{\cO}{{\cal O}}

\newcommand{\cR}{{\cal R}}
\newcommand{\cB}{{\cal B}}
\newcommand{\cC}{{\cal C}}

\newcommand{\p}{\partial}
\newcommand{\na}{\nabla}

\newcommand{\ep}{\epsilon}

\newcommand{\talpha}{{\tilde \alpha}}

\title{\boldmath 
Condensate flow in holographic models in the presence of dark matter}

\author[1]{Marek Rogatko\note{rogat@kft.umcs.lublin.pl}}
\author[2]{Karol I. Wysokinski\note{karol.wysokinski@poczta.umcs.lublin.pl}}
\affiliation{Institute of Physics \\
Maria Curie-Sk{\l}odowska University \\
20-031 Lublin, pl. Marii Curie-Sk{\l}odowskiej 1, Poland}

\emailAdd{rogat@kft.umcs.lublin.pl}
\emailAdd{karol@tytan.umcs.lublin.pl}

\abstract{
Holographic model of a three-dimensional current carrying superconductor 
or superfluid with {\it dark matter} sector 
described by the additional $U(1)$-gauge field 
coupled to the ordinary Maxwell one, has been studied in the probe limit. 
We investigated 
analytically by the Sturm-Liouville variational method, the holographic s-wave and p-wave 
models in the background of the AdS soliton as well as five-dimensional AdS black hole spacetimes. 
The two models of p-wave superfluids were considered, the so called $SU(2)$ and the Maxwell-vector.
Special attention has been paid to the dependence of the critical chemical potential
and critical transition temperature on the velocity of the condensate and {\it dark matter}
parameters.  
The current $J$ in holographic three-dimensional superconductor studied here, shows the linear
dependence on $T_c-T$ for both s and p-wave symmetry. This is in a significant contrast 
with the previously obtained results for two-dimensional superconductors, which reveal the $(T-T_c)^{3/2}$
temperature dependence.    
The coupling constant $\alpha$, as well as, chemical potential $\mu_D$ and the velocity $S_D$ of  
the {\it dark matter}, affect the critical chemical potential of the p-wave holographic $SU(2)$
system. On the other hand,  $\alpha$,  {\it dark matter} velocity $S_D$ and density $\rho_D$
determine the actual value of the transition temperature of the same
superconductor/superfluid set up. However, the {\it dark matter} does not affect the value of 
the current. }


\begin{document} 
\maketitle
\flushbottom

\section{Introduction}
\label{sec:intro}
The application of gauge/gravity duality \cite{mal98}-\cite{sac12} to study {\it experimentally} 
relevant systems has resulted in number of important findings~\cite{erdbook,zaanen-book}. Among the most intriguing 
issues is the universal ratio between the viscosity and entropy density in the strongly interacting 
system, which has helped to understand the viscosity of the quark-gluon plasma.
Other equally important findings are related to strongly coupled condensed matter systems 
and their transport properties. They include analysis of holographic superconductors, superfluids 
and their behavior under special conditions. Based on the AdS/CFT correspondence the 
description of holographic s-wave superconductor was presented \cite{gub08}-\cite{mae08}
and soon after the method was adopted to take into account p \cite{gub08} -\cite{liu15} 
and d-wave superconductors \cite{che10}-\cite{zen10}.
In \cite{amo14} a generalization of the standard holographic p-wave superconductor 
with two interacting vector order parameters was investigated. The model was
proposed as a holographic effective theory of  a strongly coupled ferromagnetic superconductor.

The aforementioned studies were generalized in many ways, e.g., back-reaction of the
 order parameter on spacetime metric was considered enabling the second order phase 
transition  to be replaced
by the first order one \cite{amm10p}, the gravitational background of AdS soliton \cite{hor98} 
was proposed to study holographic insulator/superconductor  transition at zero 
temperature \cite{nis10}-\cite{akh11}.
Further, the studies concerning Gauss-Bonnet gravity, non-linear electrodynamics 
and Weyl corrections on p-wave holographic phase transitions were 
elaborated  \cite{pan11}-\cite{jin12}, 
On the other hand, holographic vortices and droplets influenced by 
magnetic field were discussed in \cite{alb08}-\cite{roy12}.

The early applications of the gravity methods
to study holographic superconductors described by the scalar field 
in the appropriate  gravity background have soon been extended to other symmetries
and systems with super-current.

Properties of current carrying superconductors have recently been studied by
means of AdS/CFT correspondence with the hope to unveil the novel strong coupling properties of
superconductors with a constant velocity or 
super-current~\cite{gub08a,hor08,her09a,bas09,son10,are10,zen11,zen13,wu14,amm10,ama13,ama14,lai16,bas09a}. 
In the gauge/gravity duality the super-current 
is introduced by the spatial component of the gauge field depending on the radial direction.
Close to the boundary of the spacetime the constant part is interpreted as superfluid velocity, while
the component in front of the next leading term ($\propto 1/r^2$) is related 
to the current in the dual field theory.
Continuation of the work \cite{hor08}, constitutes the paper \cite{her09a} 
in which the authors consider superfluid with superfluid current flowing 
through the system. It was established that
there was a first order phase transition between superfluid and the normal 
phase in the case when one changed the superfluid current. At high temperatures
 the phase transition became a second order.
In \cite{bas09}, in the background of four-dimensional AdS planar black hole, 
the possibility of a DC-current existence was considered. For the purpose 
in question, both time and spatial components
of Maxwell potential were turned on. The critical point, where the second
 order superconducting phase transition changed to a first one, was envisaged.

It was shown \cite{son10} that in the strongly backreacted regime at low 
value of the charge, the phase transition remains of the second order one. The
direct studies of superconducting film, reveal that a DC current affects 
the superconducting phase transition, making it first order one for any non-vanishing
value of the current \cite{are10}. Other studies of $p$ and $p+ip$-wave
 holographic superconductors with fixed super-current reveal that close 
to the critical temperature,
the critical current is proportional to $(T_c-T)^{3/2}$ 
and the phase transition in the presence of it is a first order one \cite{zen11}.
One remarks that studying three-dimensional current carrying holographic superconductors, 
we have obtained the linear dependence  of the current on $(T_c-T)$  which 
seems to agree with some experimental measurements, as discussed in the last section of the paper.

The studies of one-dimensional s-wave holographic superconductor with
 a super-current caused by non-zero $A_x$ component of the Maxwell potential, 
confirmed the
previously obtained results \cite{zen13}. 
Further, the studies of holographic superfluids in the probe limit, for spacetime dimensions equal to three and four and various 
values of scalar field masses were conducted in \cite{are10a}. It turned out that $T_c$ 
decreases when the super-current increases and the order of the phase transition 
changes from second to first.
For sufficiently large value of the mass it remains second order phase transition, 
independently how high the superfluid velocity is.

Numerical investigations of a holographic p-wave superfluid in four and five dimensional 
AdS space-time, in the model with complex Maxwell vector field, reveal that for the 
condensate with fixed superfluid velocity, the results are similar to s-wave case \cite{wu14}.
 Moreover, it was observed that the increase of the superfluid velocity causes 
the inversion from the second to the first order phase transition. The larger $m^2$ was, 
the larger translating superfluid velocity one received.
On the other hand, numerical studies of p-wave Maxwell vector superfluid model in the
 background of four-dimensional Lifshitz black hole were presented in \cite{wu15}.
Investigations in AdS black hole background of p-wave superfluids with back-reaction, 
were conducted in \cite{amm10}. Without 
 back-reaction, the phase transition to the superfluid state is
a second order. On the contrary, back-reaction reveals that one can find a critical 
value of the parameter which describes the ratio of 
five-dimensional gravitational constant to Yang-Mills coupling, when  the phase 
transition is a first order one.
The question of stability of holographic superfluids with finite velocity, using 
quasi-normal  mode spectrum was treated in \cite{ama13,ama14}.

Analytical studies of s and p-wave holographic superfluids, in the AdS solitonic 
background, by means of Sturm-Liouville method, were presented in \cite{lai16}. 
It was observed that in the spacetime 
under consideration the holographic superfluid phase transition is always second-order one.
One of our aims is to analyze the holographic superconductor at finite temperatures
and calculate $T$ dependence of the super-current in an analytic way. The results are compared to
recent experimental work on temperature dependence of the super-current. 
 
The other goal of the paper is related to the tantalizing and long-standing 
question in the contemporary astrophysics and particle 
physics which is the problem of {\it dark matter}  in our Universe. The latest astronomical 
observations reveals that
almost 24 percent of the matter filling Universe constitutes {\it dark matter}. 
Nowadays, cosmological measurements \cite{cos1}-\cite{cos2} enable us to determine 
the abundance of {\it dark matter} with exquisite precision. Both observations and computer 
simulations
broaden our knowledge about {\it dark matter} distribution in galactic halos. But still its
 nature and experimental detection remain a mystery.

AdS/CMT (condensed matter theory) duality is a method how to treat condensed matter problems by means
of the gravity theory. The duality is a sort of calculus which require the AdS spacetime for the mathematics
to work in a prescribed way. However, 
the content of the gravity theory defines the conditions under which the
condensed matter problem or phenomenon  is studied. For example, the presence of the black soliton in the bulk 
means that the boundary theory is analyzed at zero temperature, while  
the presence of the black hole is connected with finite temperature effects in the dual theory.
In the same spirit the composition of the matter on the gravity side gives information
about its influence on the studied phenomenon. This is the general conviction. Based on it  we are exploring here
 the effect of the {\it dark matter} on the properties of superconductors. If the duality really
has something to say about real life systems, than the hope is that the calculated effects may one day be
discovered in the laboratory.

Thus the main question of astrophysical significance posed in our research is the request how {\it dark matter}
(present in our Universe) modifies the properties of superfluids studied here in the laboratories. 
Perhaps one can find some effects which can guide the future experiments enabling the
 detection of the aforementioned elusive component of the Universe.
Studies of the influence of dark matter on various condensed matter 
systems~\cite{nak14}-\cite{pen15b} are timely and 
of great interest  in view of proposing new ways~\cite{reg15}-\cite{nak15aa} 
to detect this hidden component of the matter in our Universe.

We shall construct a holographic superfluid model with a super-flow by allowing for  
the t-component and additional, spatial component of the gauge field.
Moreover, except of Maxwell $U(1)$ field we consider the other gauge field
 representing the {\it dark matter}, coupled to the ordinary one \cite{bri09}.
The model of {\it dark matter} discussed here is supported by numerous 
astrophysical observations \cite{ger15}-\cite{massey15b}
and other experimental data  related to the muon anomalous magnetic 
moment \cite{muon}, as well as, experimental searches for 
the {\it dark photon} \cite{afa09}-\cite{red13}. 
In the paper we discuss three models of the superfluids, i.e., s-wave, p-wave Maxwell 
vector model and  $SU(2)$ one in the gravitational background of the AdS soliton.

The paper is organized as follows. In section 2 we describe s-wave model of holographic 
superfluid at zero temperature in the space-time of AdS soliton, paying attention to the
 behavior of the 
superfluid velocity near the critical point, 
the relations of condensate operators and charge to the difference between chemical
 potential and its critical value. We also study the behavior of the spatial 
component of Maxwell potential
near the critical point. In section 3 we describe two models of p-wave holographic 
superfluids with {\it dark matter} sector. Because of the fact 
that equations of motion for the Maxwell vector model
 are similar to the s-wave case, we restrict our attention 
to the $SU(2)$ model. As in the previous model we analyze critical chemical 
potential, charge density, the 
behavior of spatial component of Maxwell field and the influence  of the {\it dark matter} 
on them. 
Section 5 is devoted to the studies of s and p-wave superfluids 
in the background of five-dimensional AdS black hole, i.e., we shall 
investigate properties of the aforementioned
superfluids at a certain temperature due to the presence of black hole in the bulk.
We summarize and conclude our investigations in section 6.

\section{Holographic s-wave superfluid model in soliton background}
In this section we introduce a set up for zero temperature s-wave holographic superfluid.
Its action is given by
\be
S = \int d^5 x \sqrt{-g}~\bigg( R - 2 \Lambda \bigg) + S_m,
\ee
where $\Lambda = - 6/L^2$,~$L$ stands for the radius of the AdS space-time, while
the action for the matter fields is taken as 
\ben
S_m = \int d^5 x \sqrt{-g}~\bigg( - \frac{1}{4} F_{\mu \nu}F^{\mu \nu} 
 &-& \frac{1}{4} B_{\mu \nu} B^{\mu \nu}  - \frac{\alpha}{4} F_{\mu \nu}B^{\mu \nu} + \\ \nonumber
&-& ( \na_\mu \psi - iqA_\mu )^{\dagger} (\na_\mu \psi - iqA_\mu)  + V(\psi) \bigg),
\een
where the potential is of the form $V(\psi) = m^2 \psi^2.$ ~$F_{\mu \nu} = 2 \na_{[\mu}A_{\nu]}$ 
denotes the strength tensor for the ordinary Maxwell field, whereas
$B_{\mu \nu} = 2 \na_{[\mu}B_{\nu]}$ is responsible for the other $U(1)$-gauge 
field which represents the {\it dark matter}  sector.
$\alpha$ stands for the coupling constant between both gauge fields, $m,~q$ is
 mass and charge of the scalar field $\psi$, respectively. This model was widely 
used in the probe limit studies, as well as,  backreaction
effects were taken into account in order  to envisage the influence of the 
{\it dark matter} on the properties of holographic s and p-wave superconductors and
 vortices \cite{nak14}-\cite{pen15b}.

Assuming that $\psi$ is real and it constitutes a function of $r$-coordinate, one has 
the following equations of motion:
\ben
\na_\mu F^{\mu \nu} &+& \frac{\alpha}{2} \na_\mu B^{\mu \nu} - 2~q^2~\psi^2~A^\nu = 0,\\
 \na_\mu B^{\mu \nu} &+& \frac{\alpha}{2} \na_\mu F^{\mu \nu} = 0,\\
 \na_\mu \na^\mu \psi &-& q^2~A_\mu A^\mu~\psi - \frac{1}{2} \frac{\p V}{\p \psi} = 0.
 \een
 The first two equations can be combined to the relation of the form
 \be
 \talpha~\na_\mu F^{\mu \nu} - 2~q^2~\psi^2~A^\nu = 0,
 \label{eqmot}
 \ee
where we have denoted $\talpha = 1 - 1/4~\alpha^2$.

The gravitational background of gauge/gravity correspondence is described by the line element of the 
five-dimensional AdS soliton spacetime. It implies
\be
ds^2 = -r^2~dt^2 + L^2~\frac{dr^2}{f(r)} + f(r)~d \phi^2 + r^2~(dx^2 + dy^2),
\label{sol}
\ee
where $f(r)= r^2 - r_0^4/r^2$,~$r_0$ denotes the tip of the line element which 
constitutes a conical singularity
of the considered solution. 
In what follows, without loss of generality,  one sets the radius of the AdS 
space-time $L$ equal to one.
The AdS solitonic solution  may be achieved from the five-dimensional 
Schwarzschild-AdS black hole spacetime
by implementing two Wick rotations. 
The Scherk-Schwarz transformation of 
$\phi$-coordinate in the form $\phi \sim \phi + \pi/r_0$,
enables to get rid of this inconvenient feature of the gravity background. The temperature
of the aforementioned background equals to zero.
In the AdS/CMT correspondence, the gravitation background in question provides a description 
of a three-dimensional field theory with a mass gap, resembling 
an insulator in condensed matter physics.

In what follows we assume that the components of the Maxwell gauge field are given by
$A_{t}(r) = \varphi(r)$  and $A_\phi (r)$. On this account, the equations of motion are provided by
\ben \label{meff}
\psi'' &+& \bigg( \frac{f'}{f} + \frac{3}{r} \bigg) ~\psi' - \frac{1}{f}~\bigg(
m^2 + \frac{q^2~A_\phi^2}{f} - \frac{q^2~\phi^2}{r^2} \bigg) \psi = 0,
\label{s-psi-1} \\
\varphi'' &+& \bigg( \frac{f'}{f} + \frac{1}{r} \bigg)~\varphi' -  \frac{2~q^2~\psi^2}{\talpha~f}~\varphi = 0,
\label{s-varphi-1}\\
A_\phi'' &+& \frac{3}{r}~A_\phi' -  \frac{2~q^2~\psi^2}{\talpha~f} ~A_\phi = 0,
\label{s-aphi-1}
\een
where the prime denotes derivative with respect to $r$-coordinate.
Let us note that in the absence of {\it dark matter} ( $\alpha=0$ and $\tilde{\alpha}=1$) the 
above equations reduce to those describing the s-wave superconductor in a model with no {\it dark matter}
sector~\cite{lai16}. Moreover, the existence of the condensate $\psi^2\ne 0$ couples 
otherwise independent components of the gauge field $A_t(r)$ and $A_\phi(r)$. 
The coupling among condensing $\psi$ and the aforementioned components of gauge field, may cause the black object (black soliton or black hole) to be unstable
to forming scalar hair. The effective mass of scalar field $\psi$ is given by $m^2_{eff}=m^2+q^2g^{tt}A_t^2+q^2g^{\phi \phi}A_\phi^2$ (see the equation \eqref{meff}).
The term proportional to $g^{tt}$ may become sufficiently negative near the event horizon to destabilize the scalar field,  as explained~\cite{gub08,hor11} for a model with $A_\phi =0$. 
The existence of the {\it dark matter} sector
modifies the underlying fields, effectively  resulting in the replacement 
of $\psi(r)$ by $\psi(r)/\tilde{\alpha}$, but does not effect the physics of the transition.
This is due to our minimal coupling between visible and {\it dark matter } sectors.  
It should be remarked that for $\psi\equiv 0$ there is no coupling among various fields and both components of the gauge field
 become  mutually independent.

In order to solve the above equations one should impose the adequate boundary conditions on 
the tip of the AdS soliton and at infinity. 
Let us remark that the form of equation (\ref{eqmot}) is such that the influence of the 
{\it dark matter} sector shows up  by appearing $\alpha$-coupling constant in the relations
governing the ordinary Maxwell field. Nevertheless, for the completeness of our 
considerations we take into account the required behavior of {\it dark matter} fields.
The fields in the considered theory are supposed to behave as
\ben
\psi &=& \psi_0 + \psi_1 (r-r_0) + \psi_2 (r-r_0)^2 + \dots,\\
\varphi &=& \varphi_{(0)} + \varphi_{(1)} (r-r_0) + \varphi_{(2)} (r-r_0)^2 + \dots,\\
\label{bc-afi} 
A_\phi &=& A_{\phi (0)} + A_{\phi (1)} (r-r_0) + A_{\phi (2)} (r-r_0)^2 
+ \dots,\\
B_t &=& B_{t (0)} + B_{ (1)} (r-r_0) + B_{t (2)} (r-r_0)^2 + \dots,\\
B_\phi &=& B_{\phi (0)} + B_{\phi (1)} (r-r_0) + B_{\phi (2)} (r-r_0)^2 + \dots,
\een
where $\psi_a,~\varphi_{(a)}, ~A_{\phi (a)}~B_{t(a)},~B_{\phi (a)}$, for $a = 0,~1,~2,\dots $
 are the appropriate integration constants. 
In order to obtain the finiteness of the above quantities the Neumann-like boundary conditions
are required to be satisfied. Let us remark that the Neumann-boundary conditions 
were widely treated, e.g., in \cite{dom10} where
the dynamical gauge fields subject to the aforementioned conditions on the AdS boundary were implemented.

At the tip of the the considered AdS soliton, one requires that $\phi, ~B_t$ 
will have constant non-zero value (contrary to the behavior at the black hole event horizon,
where the quantities in question are equal to zero).
On the other hand, at the asymptotic AdS boundary, when $r \rightarrow \infty$, one has to satisfy
the following relations:
\be
\psi = \frac{\psi_-}{r^{\Delta_-}} + \frac{\psi_+}{r^{\Delta_+}},
\qquad \varphi = \mu - \frac{\rho}{r^2}, \qquad A_\phi = S_\phi - \frac{J_\phi}{r^2},
\label{bc}
\ee
where $\Delta_\pm = 2 \pm \sqrt{4 + m^2},~\mu$ and $S_\phi$ stand for the chemical
 potential and superfluid velocity, respectively. $\rho$ is the charge
density, while $J_\phi$ gives  the current in the dual field theory.  
Both quantities $\psi_-$ and $\psi_+$ multiply normalizable modes of the scalar field equation. 
According to the AdS/CFT correspondence, they constitute the vacuum expectation
values $\psi_- = <\cO_->$ and $\psi_+ = <\cO_+>$ of the operator dual to the scalar field. 
One can impose the boundary conditions that either $\psi_-$ or $\psi_+$ vanish.
As was revealed in \cite{har08} imposing the boundary conditions in which $\psi_-$ and $\psi_+$ 
are nonzero caused the asymptotic AdS theory unstable \cite{her04,her05,sio10}.
Moreover, there are two alternative quantizations for the scalar field in $AdS_5$, i.e.,
 the operators are normalizable \cite{kle99} if 
$0<\sqrt{m^2 + 4} <1$, which implies that $-3> m^2 >-4$. 
In the following  we shall use $\psi_+\ne 0$ and let $\psi_-=0$.

As a first preparatory step, let us rewrite the equations (\ref{s-psi-1})-(\ref{s-aphi-1}) 
in the new coordinates $z=r_0/r$.
They can be rewritten in the following form
\ben
\psi'' (z) &+& \bigg( \frac{f'(z)}{f(z)} - \frac{1}{z} \bigg)~\psi'(z)  - \frac{1}{f(z)}~\bigg(
\frac{m^2}{z^4} + \frac{q^2~A_\phi^2(z)}{r_0^2~ f(z)~z^4} - 
\frac{q^2~\phi^2(z)}{r_0^2 ~z^2} \bigg) \psi(z) = 0,\\
\varphi''(z) &+& ~\bigg( \frac{f'(z)}{f(z)} + \frac{1}{z} \bigg) ~\varphi'(z) - 
 \frac{2~q^2~\psi^2(z)}{r_0^2~\talpha~f(z)~z^4}~\varphi(z) = 0,\\
A_\phi''(z) &-& \frac{1}{z}~A_\phi'(z) -  
\frac{2~q^2~\psi^2(z)}{r_0^2~\talpha~f(z)~z^4} ~A_\phi(z)= 0,
\label{a-phi-2}
\een
where $f(z) = (1-z^4)/z^2$.

In the next sections we solve these equations analytically close to the transition point \cite{sio10} and obtain 
the critical value of the chemical potential $\mu_c$, the behavior of the order parameter
and $A_\phi$ component of the Maxwell field.

\subsection{Critical chemical potential for s-wave superfluid}

On the gravity side, the transition we are discussing here results from   
the addition of the chemical potential $\mu$ to the soliton. The resulting 
solution is unstable towards
scalar hair for $\mu$ bigger than the critical one. The resulting system with a mass
gap is on the field theory side interpreted as an insulator. At zero
 temperature and for $\mu>\mu_c$ the system
undergoes phase transition to the superfluid phase.

At the critical potential $\mu_c$, the order parameter $\psi \sim 0$ 
and the equation of motion for $\varphi$-gauge field implies
\be
\varphi''(z) + \bigg( \frac{1}{z} + \frac{f'}{f} \bigg)~\varphi'(z)  \simeq 0,
\ee
then $\varphi = \mu + c_1~\log (1+z^2)/(1-z^2)$. The boundary conditions at
the tip of the soliton require that $c_1=0$. 
It leads to the condition that $\varphi(r)$ has the constant value $\mu$, when
the order parameter tends to zero.
Under the same conditions the $A_\phi$-component fulfills the equation
\be
A_\phi''(z) - \frac{1}{z} A'_\phi(z) \simeq 0,
\label{aph}
\ee
with a solution  $A_\phi = S_\phi (1 - z^2)$. 
It fulfills the boundary conditions $A_\phi (1) = 0$ as required in \cite{lai16}.
When $\mu \rightarrow \mu_c$ one gets the following equation for $\psi$ 
\be
\psi''(z) + \bigg( \frac{f'}{f} - \frac{1}{z} \bigg) \psi'(z) - \frac{1}{f}~\bigg(
\frac{m^2}{z^4} + \frac{q^2~S_\phi^2(1-z^2)^2}{r_0^2 ~f~z^4} - \frac{q^2~\mu^2}{r_0^2~ z^2} \bigg) \psi(z) = 0.
\label{p}
\ee
The boundary conditions for the equation (\ref{p}) are given by the relations  (\ref{bc}).
In order to get information valid close to the boundary ($z \rightarrow 0$), where the
field theory lives, we suppose that $\psi$ can be approximated by
\be
\psi \sim <\cO_+>~z^{\Delta_+}~F(z),
\ee
where  $\Delta_+ = 2 + \sqrt{4 + m^2}$. 
For the function $F(z)$ we impose the standard boundary conditions $F(0)=1,~F'(0)=0$ and
for numerical illustration take $F(z)=1-az^2$, with $a$ being a variational parameter.

In principle, the are two slightly different ways to treat the problem in question. 
Both rely on the Sturm-Liouville variational method \cite{sio10}. 
In the first one, used in \cite{lai16},  one introduces a dimensionless parameter
\be
k = \frac{S_\phi}{\mu_c},
\ee
and rewrites the relation (\ref{p})  as 
\ben
F''(z) &+& F'(z) \bigg[ \frac{2 \Delta_i}{z}  + \bigg( \frac{f'}{f} - 
\frac{1}{z} \bigg) \bigg] \\ \nonumber
&+& F(z)~\bigg[
\frac{\Delta_i~(\Delta_i -1)}{z^2} + \bigg( \frac{f'}{f}-\frac{1}{z} \bigg)\frac{\Delta_i}{z} - 
\frac{1}{z^2~f}~\bigg( \frac{m^2+ (1-z^2)^2~q^2~S_\phi^2/r_0^2}{z^2} 
- \frac{\mu_c^2~q^2}{r_0^2} \bigg) \bigg] = 0.
\label{ff}
\een
The above equation can easily be converted into standard  Sturm-Liouville one
\be
(p(z)~F'(z))' - q(z)~F(z) + \la^2~r(z)~F(z) = 0,
\label{SL1}
\ee
where we have denoted
\ben
p(z) &=& z^{2 \Delta_i -1}~f(z),\\
q(z) &=& - z^{2 \Delta_i -1}~f(z)~\bigg[ \frac{\Delta_i (\Delta_i -1)}{z^2} + \bigg(
\frac{f'}{f} - \frac{1}{z} \bigg)~\frac{\Delta_i}{z} - \frac{m^2}{z^4~f} \bigg],\\
r(z) &=& z^{2 \Delta_i -3}~\bigg(
1 - \frac{(1-z^2)^2}{z^2}~k^2 \bigg).
\een
The Sturm-Liouville eigenvalue problem enables us to study a variational  
determination of the  lowest eigenvalue $\lambda^2 = \mu^2~q^2/r_0^2$.
Varying the following functional we estimate the lowest value of the aforementioned spectral parameter
\be
\lambda^2 = \frac{\mu^2~q^2}{r_0^2} = \frac{ \int_0^1dz~ [ F'(z)^2~p(z) +
 q(z)~F^2(z)]}{\int_0^1 dz~r(z)~F^2 (z)}.
\label{fntl1}
\ee
The above Sturm-Liouville problem is subject to the divergent behavior for 
large values of $k^2$, due to
the structure of $r(z)$ function in the denominator.

To avoid such unphysical divergences we
rewrite the Sturm-Liouville equation (\ref{SL1})  
with the following functions ${\tilde p}(z),~{\tilde q}(z)$ and ${\tilde r}(z)$ 
and the new parameter $K=q~S_\phi/r_0$. They yield
\ben
{\tilde p}(z) &=& z^{2 \Delta_i -1}~f(z),\\
{\tilde q}(z) &=& - z^{2 \Delta_i -1}~f(z)~\bigg[ \frac{\Delta_i (\Delta_i -1)}{z^2} + \bigg(
\frac{f'}{f} - \frac{1}{z} \bigg)~\frac{\Delta_i}{z} - \frac{m^2+K^2(1-z^2)^2}{z^4~f} \bigg],\\
{\tilde r}(z) &=& z^{2 \Delta_i -3}.
\een
Again, the Sturm-Liouville eigenvalue problem enables us to  
determine  $\Lambda^2 = \mu_c^2~q^2/r_0^2$ as a spectral parameter and estimate 
its minimum eigenvalue by the variation of the functional given by
\be
\Lambda^2 = \frac{\mu_c^2~q^2}{r_0^2} = \frac{ \int_0^1dz~ [ F'(z)^2~{\tilde p}(z) +
 {\tilde q}(z)~F^2(z)]}{\int_0^1 dz~{\tilde r}(z)~F^2 (z)}.
\label{fntl2}
\ee
This form of equation is free of the unphysical divergences for large value of $k$. It
has to be noted that, while $k$ formally is a function of $\mu_c$ the parameter $K$ does 
not depend on $\mu$.
It enters the function ${\tilde q}(z)$ and leads to continuous increase of $\la$ with $K$.

\begin{figure}
\includegraphics[width=0.49\linewidth]{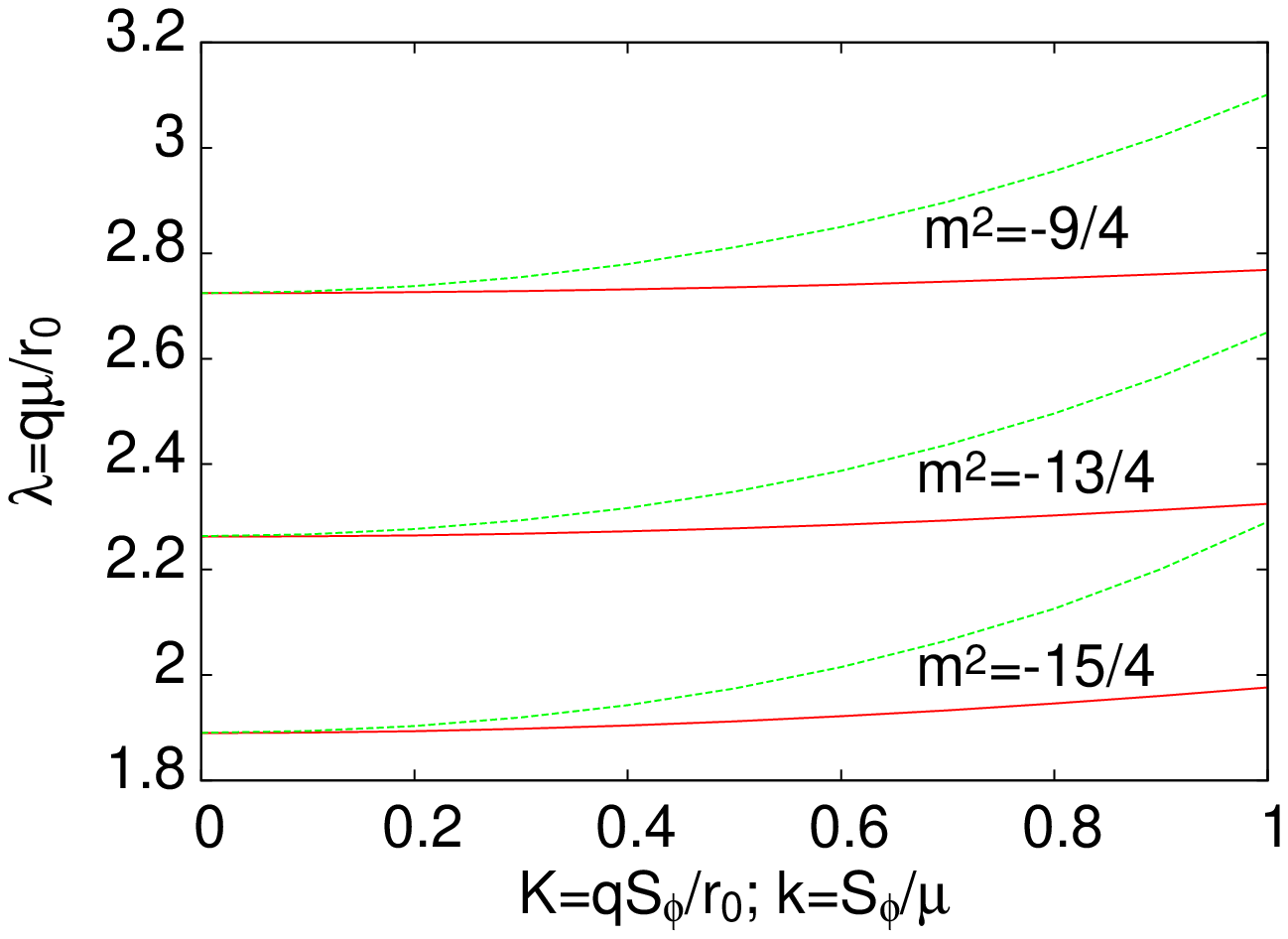}
\includegraphics[width=0.49\linewidth]{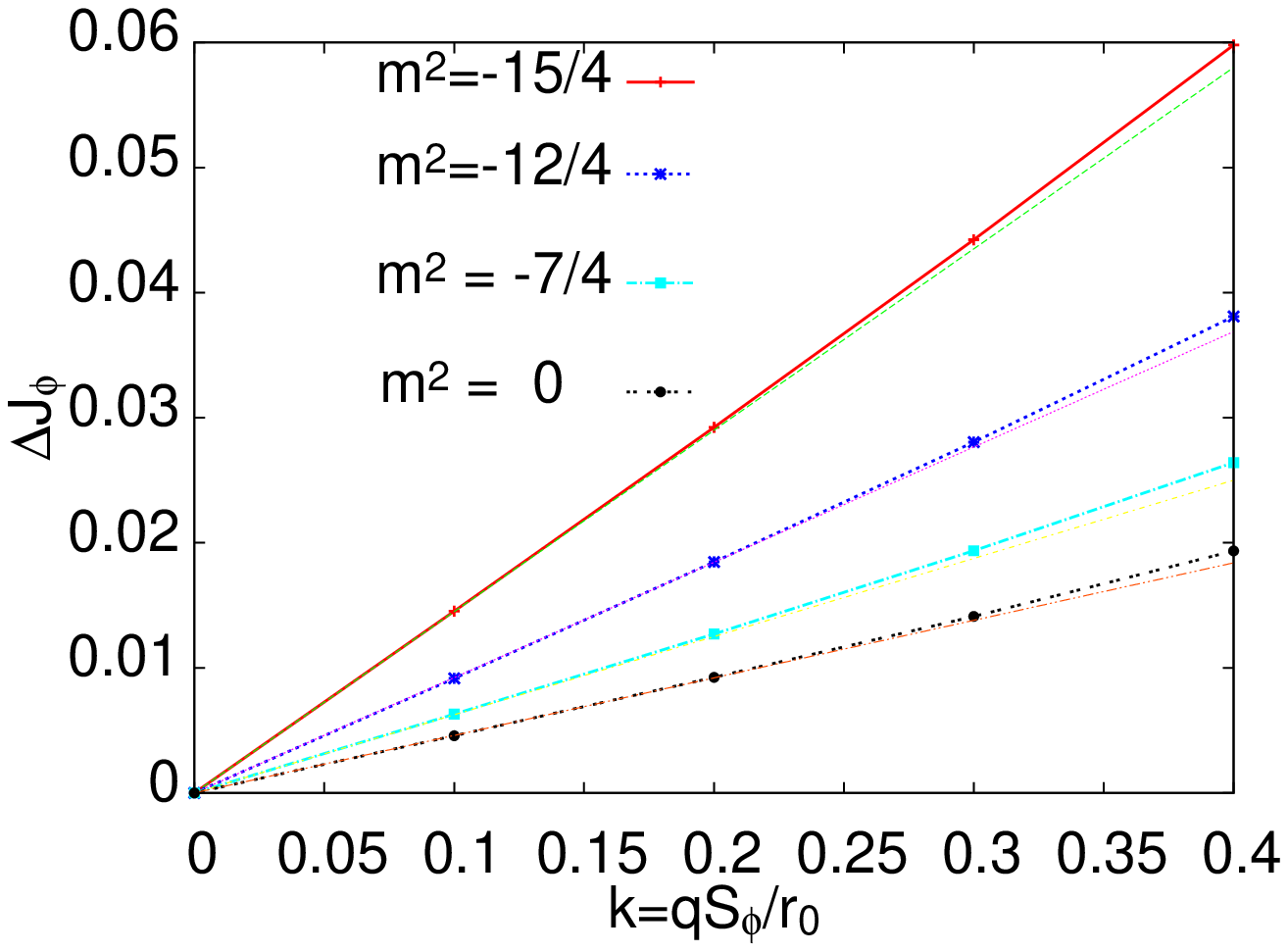}
\caption{(left panel) The dependence of the critical chemical potential on the superflow
parameter $S_\phi$. We plot $\lambda=q\mu/r_0$ {\it vs.} $K=qS_\phi/r_0$ 
(green, dashed curves)
or the same spectral parameter {\it vs.} $k=S_\phi/\mu$ (red,  solid curves),
for $m^2$ equal to $-9/4,~-13/4,~-15/4$, respectively. One observes 
quadratic increase of $\lambda$ with $k$ and $K$, albeit on different scales. 
(right panel) The dependence of the super-current
 $\Delta J_\phi=(J_\phi-S_\phi)/(q^2~S_\phi~<\cO_i>^2)$  in a s-wave
superconductor on the velocity $S_\phi$,  for a few values of the
 parameter $-4=m^2_{BF} < m^2\le 0$ (curves with symbols). Curves
without symbols correspond to linear approximations for low values of $S_\phi$.
}
\label{fig1}
\end{figure}

For numerical purposes we have set $q=1$ and $r_0=1$.
Figure \ref{fig1} (left panel) illustrates the dependence on the critical value of the chemical potential
 in both cases, or more
precisely, $\la$ ($\Lambda$) as a function of 
$k$ ($K$). We show both dependencies on the same plot, but one has to take into account 
that the units of $k$ are different from $K$. This means that the actual values of
 $S_\phi$ are much bigger if parametrized by $k$. 
For both ways of calculations, the parameter $\lambda (\Lambda)$ increases in an approximately 
quadratic manner with  $S_\phi$, starting with the same value 
for $S_\phi=0$. However, this dependence seems to be much slower if $S_\phi$ is measured 
in units of the
critical potential $\mu_c$. This is related to the fact that $\mu_c$ is generally bigger 
than 1 and the actual superflow values differ. Needless to say, 
the numerical results
for $\lambda$ exactly agree with those reported in \cite{lai16}, in the appropriate cases.

\subsection{Critical phenomena}
In this subsection we shall concentrate on the relation connecting charge density 
and the chemical potential. 
Because of the fact that the equations describing 
the problem in question are the same as studied in \cite{nak15a}, we refer the
 reader to this reference for the details. 

Among all,  one finds that the order parameter $<\cO_i>$ depends on the chemical potential as 
\be
<\cO_i> = \sqrt{\frac{\talpha~(\mu-\mu_c)}{2~\mu_c~\xi(0)}}.
\label{op-2}
\ee
It shows that the critical phenomenon represents the second order phase 
transition for which the exponent has the mean field value $1/2$.
The final result describing the dependence of $\rho$ on the chemical potential $\mu$ reads
\be
\rho = \frac{\mu - \mu_c}{2~\xi(0)}~q^2~\int_0^1 dz~z^{2 \Delta_i-3}~F^2(z).
\ee
As in \cite{nak15} the charge density for s-wave holographic 
superfluid is proportional to the difference 
of $(\mu -\mu_c)$ and is independent of $\alpha$-coupling constant of {\it dark matter} sector.

\subsection{Behavior of $A_\phi$ near critical point}

To extract the relation between the super-current  $J_\phi$ and velocity  $S_\phi$ one needs to consider the asymptotic behavior of $A_\phi$.
Thus we investigate here the properties of $A_\phi$-component of Maxwell field near 
the critical point. The inspection  of the  equation of motion (\ref{a-phi-2}) reveals
\be
A_\phi'' - \frac{1}{z} A_\phi' - 
\frac{2~q^2~z^{2 \Delta_i}~<\cO_i>^2 ~r_0^2~F^2(z)}{\talpha~f(z)~z^4}~A_\phi = 0,
\ee
Then, expanding $A_\phi$ near the critical point in series, implies
\be
A_\phi \simeq S_\phi (1-z^2) +
 <\cO_i>^2~\bigg[ \kappa(0) + \kappa'(0) ~z + \frac{1}{2}~\kappa''(0)~z^2 + \dots \bigg].
\ee
The same procedure as in the preceding section implemented to $A_\phi$, enables us to find that
for $\kappa(z)$ up to 2-order in $<\cO_i>$, we get
\be
\kappa''(z) - \frac{1}{z}~\kappa(z)' \simeq \frac{2~q^2}{~r_0^2 \talpha~f(z)}
~S_\phi~(1-z^2)~z^{2\Delta_i -4} ~F^2(z).
\ee
In the next step we find that
\be
\kappa''(0) = \frac{\kappa'(z)}{z} \mid_{z \rightarrow 0} = 
- \frac{2~q^2~S_\phi}{~r_0^2}~\int_0^1 dz~\frac{(1-z^2)}{\talpha~f(z)}~z^{2 \Delta_i-5}~F^2(z),
\ee
where we have used the fact that inspection of the z-order terms reveals that $\kappa'(0) = 0$.

All the above help us to determine that
\be
A_\phi = S_\phi (1-z^2) - \frac{q^2~S_\phi~<\cO_i>^2}
{~r_0^2\talpha}~z^2~\int_0^1 dx~\frac{(1-x^2)}{f(x)}~x^{2 \Delta_i -5}~F^2(x).
\ee
The obtained relation is in accord with our boundary condition demand that $A_\phi(1) = 0$, 
at the critical point. 
The $A_\phi$ is $\alpha$-coupling dependent. The bigger
$\alpha$ one considers, the smaller value of the spatial component of Maxwell field we gain. 
Thus, the existence 
of {\it dark matter} sector diminishes the value of $A_\phi$,
causing the increase of the superfluid current $J_\phi$. 
Using the asymptotic behavior of $A_\phi$ as given in (\ref{bc}), 
the current $J_\phi$ is provided by
\be
J_\phi = S_\phi  +
 \frac{q^2~S_\phi~<\cO_i>^2}{r_0^2~\talpha}~\int_0^1 dx~\frac{(1-x^2)}{f(x)}~x^{2 \Delta_i -5}~F^2(x).
\label{curr-sfi}
\ee  
The above equation shows that the current $J_\phi$ is linearly related to
the velocity $S_\phi$. Small deviations are expected for the dependence of the integral on  $S_\phi$.
In the right panel of figure \ref{fig1}, the dependence of 
$\Delta J_\phi=(J_\phi-S_\phi)/(q^2~S_\phi~<\cO_i>^2)$ on $S_\phi$ is shown
 for a s-wave superconductor and for a few values of the parameter
 $m^2_{BF}< m^2\le 0$ and the  coupling to the {\it dark matter} $\alpha=0$. 
Only the slight deviation from the linear dependence can be observed. It is traced 
back to the neglecting of backreaction effects. 
Due to independence of $<\cO_i>^2/\talpha$ on $\alpha$ one notes that $J_\phi$
does not depend on  {\it dark matter}.

\section{Holographic p-wave superfluid model in soliton background}
\subsection{Vector model of p-wave superfluid}
 In this subsection we shall analyze the so-called Maxwell vector model of p-wave
 superfluids \cite{cai13} with {\it dark matter} sector. 
Its action is similar to the action appearing in quantum electrodynamical $\rho$-meson 
description \cite{dju05}.

The gravitational part of the model in question is the same as before, whereas the matter 
action is given by 
\ben
\label{s_matter1}
S_{m} = \int \sqrt{-g}~ d^5x  \bigg( 
- \frac{1}{4}F_{\mu \nu} F^{\mu \nu } &-& 
\frac{1}{4}B_{\mu \nu} B^{\mu \nu } - \frac{\alpha}{4}~B_{\mu \nu} F^{\mu \nu } \\ \nonumber
- \frac{1}{2}~\rho^{\dagger}_{\mu \nu}~\rho ^{\mu \nu} 
&-& m^2~\rho^{\dagger}_\mu~\rho^{\mu} + i~q~\gamma_0~\rho_\mu~\rho^{\dagger}_\nu~F^{\mu \nu}
\bigg),
\een
where $\rho_\mu$ is a complex vector field with mass $m$ and the charge $q$.
The quantity $\rho_{\mu \nu}$ is defined by means of the co-variant 
derivative $D_\mu = \na_\mu - iqA_\mu$ in the form 
$\rho_{\mu \nu} = D_\mu \rho_\nu -  D_\nu \rho_\mu.$
The last term in equation (\ref{s_matter1}) is devoted to the magnetic moment of the
 vector field $\rho_\mu$. 
In the model under consideration, a charged $U(1)$ vector field is equivalent on the AdS/CFT side
to an operator carrying the same charge under the symmetry in question. On the other hand, 
a vacuum expectation 
value of this operator is being subject to the spontaneous
$U(1)$ symmetry breaking. The condensate of the dual operator leads to 
the $U(1)$ symmetry breaking. Due to vector character of the field, 
the rotational symmetry is broken by choosing a specific spatial direction.
Therefore one can conclude that the vector field may be regarded as an order parameter 
and the model may describe p-wave superfluids.

We suppose that the vector field  is  real with only one component and 
the other gauge fields are chosen as 
\be
\rho_\alpha~dx^\alpha = \rho_x~dx, \qquad
A_\mu~dx^\mu = \varphi(r)~dt + A_\phi(r)~d\phi, \qquad 
B_\nu~dx^\nu = \eta(r)~dt + \xi(r)~d \phi.
\ee
The close inspection of the equations of motion for the underlying theory with {\it dark matter} 
sector and with the real components of the vector field, gives us the same description 
as the s-wave model 
studied before, when we exchange $\psi(r)$ (which acts as an order
parameter in s-wave case) for the $\rho_x$-component of the vector field. The same 
situation has been encountered earlier 
in the description of p-wave superconductors in \cite{rog16}.
Therefore we shall not discuss this model and concentrate on the other holographic 
superfluid p-wave model, the $SU(2)$ one.

\subsection{SU(2) Yang-Mills p-wave holographic superfluid model with dark matter sector}
This section is devoted to the basic features of the SU(2) Yang-Mills holographic p-wave
superfluid model with the {\it dark matter} sector.  The action for the matter field is given by
\be
\label{s_matter}
S_{m} = \int \sqrt{-g}~ d^5x  \bigg( 
- \frac{1}{4}F_{\mu \nu}{}{}^{(a)} F^{\mu \nu (a)} - 
\frac{1}{4}B_{\mu \nu}{}{}^{(a)} B^{\mu \nu (a)} - \frac{\alpha}{4}~B_{\mu \nu}{}{}^{(a)}F^{\mu \nu (a)} 
\bigg),
\ee
where $F_{\mu \nu}{}{}^{(a)}$ and $B_{\mu \nu}{}{}^{(a)}$ are two $SU(2)$ Yang-Mills 
field strengths of the form
$F_{\mu \nu}{}{}^{(a)} = \na_\mu A^{(a)}_\nu - \na_\nu A^{(a)}_\mu + \ep^{abc}~A^{(b)}_\mu A^{(c)}_\nu$.
The totally antisymmetric tensor is set as $\ep^{123} =1$. The components 
of the gauge fields are bounded with
the three generators $\tau$ of the $SU(2)$ algebra by the relations 
$A =  A^{(a)}_\beta~\tau^a~dx^\beta$, where
$[\tau^a,~\tau^b] = \ep^{abc}~\tau^c$. As before, the parameter $\alpha$ describes 
the coupling between ordinary and dark matter $U(1)$-gauge fields.

The equations of motion for $B_{\mu \nu}$ are provided by
\be
\na_\mu B^{\mu \nu (a)} + \frac{\alpha}{2}\na_\mu  F^{\mu \nu (a)} +
 \ep^{abc}~B_\mu{}{}^{(b)}~ B^{\mu \nu (c)}
+ \frac{\alpha}{2} \ep^{abc}~B_\mu{}{}^{(b)}~ F^{\mu \nu (c)} = 0.
\label{e1}
\ee
and for $F_{\mu \nu}$  by
\be
\na_\mu F^{\mu \nu (a)} + \frac{\alpha}{2}\na_\mu  B^{\mu \nu (a)} +
 \ep^{abc}~A_\mu{}{}^{(b)}~ F^{\mu \nu (c)}
+ \frac{\alpha}{2}~\ep^{abc}~A_\mu{}{}^{(b)}~ B^{\mu \nu (c)} = 0.
\label{e2}
\ee
In order to simplify the above equations we multiply relation (\ref{e1}) by $\alpha/2$ and extract 
the term $\frac{\alpha}{2} \na_\mu B^{\mu \nu (a)}$. The second term in the equation (\ref{e2})
is replaced by the aforementioned outcome. The final result may be written as
\ben \label{e3}
\tilde \alpha ~\na_\mu F^{\mu \nu (a)} &-& 
\frac{\alpha}{2}
\ep^{abc}~B_\mu{}{}^{(b)}~ B^{\mu \nu (c)}
- \frac{\alpha^2}{4}\ep^{abc}~B_\mu{}{}^{(b)}~ F^{\mu \nu (c)} \\ \nonumber
&+& \ep^{abc}~A_\mu{}{}^{(b)}~ F^{\mu \nu (c)} +
\frac{\alpha}{2}
\ep^{abc}~A_\mu{}{}^{(b)}~ B^{\mu \nu (c)} = 0,
\een
where $\tilde{\alpha} = 1 - \frac{\alpha^2}{4}$.

Both $A_\mu{}{}^{(b)}$ and $B_\mu{}{}^{(b)}$ fields, are dual to some current operators 
in the four-dimensional boundary field theory. 
We choose the following components of the underlying gauge fields
\ben  \label{ch}
A &=& \varphi(r)~\tau^3~dt + A_{\phi}(r)~\tau^3~d\phi +w(r)~\tau^1~dx, \\ \label{ch22}
 B &=& \eta(r)~\tau^3~dt  + \xi(r)~\tau^3~d \phi.
\een
In the above relations the $U(1)$ subgroups of $SU(2)$ group generated by $\tau^3$ are 
identified with the electromagnetic $U(1)$-gauge field ($\varphi(r)$) and the other
 $U(1)$ group connected with
the {\it dark matter} sector field ($\eta(r)$) coupled to the Maxwell one. 
The gauge boson field ($w(r)$) 
having the nonzero component  along $x$-direction is charged under $A_t^{(3)} = \varphi(r)$.
According to the AdS/CFT dictionary, $\phi(r)$ is dual to the chemical potential on the boundary, 
whereas $w(r)$ is dual  to $x$-component of a charged vector operator.
The condensation of $w(r)$ field will spontaneously break the $U(1)$ symmetry and is subject 
to the superconducting phase transition.  It breaks the rotational symmetry 
by making the $x$-direction  a special one
and inclines the phase transition. The transition in question 
is interpreted as a p-wave superconducting 
phase transition on the boundary. As far as the $U(1)$-gauge field  
related to the {\it dark matter} sector 
is concerned, it has the component $B_t^{(3)}$  dual to a current operator on the boundary.
One can remark that the choice described by the relation (\ref{ch}) is the only consistent choice 
of the gauge field components allowing the analytic treatment of the problem~\cite{rog16}.

The direct calculations reveal that 
the $x(1)$,~$\varphi(3)$ and $t(3)$ components of the equation (\ref{e1}) are given as follows 
\ben \label{eq39}
\frac{\alpha}{2}~\na_\mu F^{\mu x (1)} &+& \frac{\alpha}{2}~\bigg(
\ep^{132}~B_t{}{}^{(3)}~F^{tx (2)}  + \ep^{132}~B_\varphi{}{}^{(3)}~F^{\phi x (2)} \bigg)= 0,\\ \label{xi}
\na_{\mu} B^{\mu \phi (3)} &+& \frac{\alpha}{2}~\na_\nu F^{\nu \phi (3)} = 0, \\  \label{eta}
\na_{\mu} B^{\mu t (3)} &+& \frac{\alpha}{2}~\na_\nu F^{\nu t (3)} = 0,
\een
while the same components of the equation (\ref{e2}) yield
\ben \label{eq312}
\na_\mu F^{\mu x (1)} &+& \ep^{1 3 2}~A_\phi {}{}^{(3)}~F^{\phi x (2)} +
 \ep^{1 3 2}~A_t {}{}^{(3)}~F^{t x (2)} = 0,\\
\na_\nu F^{\nu \phi (3)} &+& \frac{\alpha}{2}~\na_\nu B^{\nu \phi (3)} +
 \ep^{3 1 2}~A_x {}{}^{(1)}~F^{x t (2)} = 0,\\
\na_\nu F^{\nu t (3)} &+& \frac{\alpha}{2}~\na_\nu B^{\nu t (3)} +
 \ep^{3 1 2}~A_x{}{}^{(1)}~F^{\mu t (2)} = 0,
\een
and the main relation  (\ref{e3}) reduces to the following:
\ben
\tilde \alpha ~\na_\mu F^{\mu x (1)} &-& \frac{\alpha^2}{4}~\ep^{1 b c}~B_\mu^{}{}^{(b)}~F^{\mu x (c)} + 
\ep^{1 b c} ~A_\mu{}{}^{(b)}~F^{\mu x (c)} = 0,\\
\tilde \alpha ~\na_\mu F^{\mu \phi (3)} &+& \ep^{3 b c}~A_\mu{}{}^{(b)}~F^{\mu \phi (c)} = 0,\\
\tilde \alpha ~\na_\mu F^{\mu t (3)} &+& \ep^{3 b c}~A_\mu{}{}^{(b)}~F^{\mu t (c)} = 0.
\een
In what follows, we rewrite the above general equations for the adequate components of $A_\mu$ and $B_\mu$ gauge fields, given by the relations  (\ref{ch}) and (\ref{ch22}),
discuss the appropriate boundary conditions, the asymptotic behavior and the relations between them 
in the superconducting and normal states.

\subsubsection{Critical chemical potentials}

As in s-wave case we shall consider the gravitational background of five-dimensional AdS soliton
characterized by the line element (\ref{sol}) 
\be
ds^2 = -r^2~dt^2 + L^2~\frac{dr^2}{f(r)} + f(r)~d \phi^2 + r^2~(dx^2 + dy^2),
\label{sol3}
\ee
where $f(r)= r^2 - r_0^4/r^2$. Let us recall that $r_0$ denotes the tip of the line
 element which constitutes a conical singularity
of the considered solution. 
Solitonic background means that we shall consider zero 
temperature model of p-wave superfluid.
In order to solve the underlying equations of motion for the p-wave holographic superfluid
model, one imposes the adequate Neuman-like boundary conditions. At the tip one has the same
boundary conditions as in the s-wave superfluid problem
\ben
w &=& w_0 + w_1~(r-r_0) + w_2~(r-r_0)^2 + \dots,\\
A_\phi &=& A_{\phi (0)} + A_{\phi (1)} (r-r_0)^2 + \dots,\\
\varphi &=& \varphi_{(0)} + \varphi_{(1)}~(r-r_0) + \varphi_{(2)}~(r-r_0)^2 + \dots,\\
\eta &=& \eta_{(0)} + \eta_{(1)} (r-r_0) + \eta_{(2)}(r-r_0)^2 + \dots,\\
\xi &=& \xi_{(0)} + \xi_{(1)} (r-r_0) + \xi_{(2)}(r-r_0)^2   + \dots,
\een
where $w_j$~$A_{\phi (j)},~\varphi_{(j)},~\eta_{(j)},~\xi_{(j)}$, for $j = 0,~1,~2, \dots$ 
are  constants. One encumbers the Neumann-like
boundary condition to obtain every physical quantity finite \cite{nis10}. 
Contrary, near the boundary
where $r \rightarrow \infty$, we have the different asymptotic behavior 
(comparing to the s-wave case).
The asymptotic solutions read 
\ben
w \rightarrow w_0 &+& \frac{w_2}{r^2}, \qquad \varphi \rightarrow \mu - 
\frac{\rho}{r^2}, \qquad A_\phi \rightarrow S_\phi - \frac{J_\phi}{r^2},\\
\eta \rightarrow \mu_D &-& \frac{\rho_D}{r^2}, \qquad B_\phi \rightarrow S_D - \frac{J_D}{r^2},
\een
where $\mu,~\mu_D$ and $\rho,~\rho_D$ are interpreted as the chemical potential and the charge density 
in the dual theory for ordinary and {\it dark matter}, respectively. 
Similarly, $S_\phi$, $J_\phi$, $S_D$ and $J_D$ are 
interpreted as velocity and current of ordinary matter superfluid and {\it dark matter} sector quantities.
Consequently with the requirements of the AdS/CFT dictionary, $w_0$ and $w_2$ have interpretations as a source 
and the expectation value of the dual operator.
In order to gain the normalizable solution, one puts $w_0 =0$ 
as we are interested in the spontaneous transitions to the condensed state.
  
In $z$-coordinates (with $r_0=1$) the equations in question yield
\ben \label{w1}
w''(z) &+& \bigg( \frac{f'(z)}{f(z)} + \frac{1}{z} \bigg)~w'(z) 
+ \frac{\varphi (z)[\varphi(z)-\frac{\alpha^2}{4}~\eta(z)]}{\tilde \alpha~
f(z)~z^2}~w(z) + \\ \nonumber
&-& \frac{A_\phi (z)[A_\phi(z)-\frac{\alpha^2}{4}~\xi(z)]}{\tilde \alpha~
f^2(z)~z^4}~w(z) = 0,\\ \label{w2}
A_\phi''(z) &-& \frac{1}{z}~A_\phi'(z) - \frac{w^2(z)}{\tilde \alpha~z^2~f(z)}~A_\phi(z)
= 0,\\
\varphi''(z) &+& \bigg( \frac{f'(z)}{f(z)} + \frac{1}{z} \bigg)~\varphi'(z) - 
\frac{w^2(z)}{\tilde \alpha~z^2~f(z)}~\varphi(z)
= 0.
\een
The obtained equations can  be benchmarked against the known 
results for SU(2) model of p-wave holographic superconductors. Putting
$\talpha=1$ and $A_\phi=0$ one obtains the set of equations studied earlier in~\cite{cai15} ,
albeit in 3+1 dimensional background.
In analogy to the discussion of s-wave case we remark again, that it is a condensation of the
field  $w(z)$ which, when non-vanishing, couples various gauge fields and makes them linearly
dependent everywhere in the bulk. On the other hand if, the vector condensate vanishes, $w(z) \equiv 0$, the
various components of the gauge fields become independent.

In the next step we find the dependence of $\varphi (z)$ and $A_\phi (z)$ on the component
 of the {\it dark matter} sector $\eta(z)$ and $\xi(z)$.
Using the the adequate components of the metric tensor for the line element (\ref{sol3}) 
and the equations (\ref{xi})-(\ref{eta}), 
 we arrive at the following relations  
\ben
\xi(z) &+& \frac{\alpha}{2} ~A_\phi = C_1~(1-z^2),\\
\eta(z) &+& \frac{\alpha}{2}~\varphi(z) = D_2,
\een
where $C_1$ and $D_2$ are integration constants. The integration constant 
$D_1$ we put equal to zero at $z=0$, due to the well behavior of the functions. We set
$D_2 = \mu_D$.  On the other hand, the relation between $C_1$ and $C_2$ was established 
taking into account the boundary conditions
$\xi(1) = 0$ and $A_\phi(1) =0$, then identify the integration constants with 
{\it dark matter} characteristics $S_D$ and $\mu_D$, one obtains
\ben
\xi(z) &=& S_D~(1-z^2) - \frac{\alpha}{2}~A_\phi(z),\\
\eta(z) &=& \mu_D - \frac{\alpha}{2}~\varphi(z).
\een
For $z$ close to boundary $z\rightarrow 0$,  we use the fact that $\varphi(z)=\mu$ 
and $A_\phi=S_\phi(1-z^2)$, which in turn leads to the following relations:
\ben \label{b1}
\xi(z) &=& S_D~(1-z^2) - \frac{\alpha}{2}~S_\phi(z)(1-z^2),\\ \label{b2}
\eta(z) &=& \mu_D - \frac{\alpha}{2}~\mu(z).
\een

Let us turn back to the problem of the consistency of the chosen ansatz. Using the equation \eqref{eq39} and \eqref{eq312} multiplied by $\alpha/2$,  as well as having in mind the relations
\eqref{b1} and \eqref{b2}, we arrive at
\be
\bigg( - \frac{(2 + \alpha)}{\alpha}~B_\phi{}{}^{(3)} + \frac{2}{\alpha} S_D~(1-z^2) \bigg)~F^{\phi x (2)} +
\bigg( - \frac{(2 + \alpha)}{\alpha}~B_t{}{}^{(3)} + \frac{2}{\alpha} \mu_D \bigg)~F^{t x (2)} = 0.
\label{rel}
\ee
Having in mind that 
\be
F_{\phi x}{}{}^{ (2)} = A_\phi{}{}^{(3 )} A_x{}{}^{(1)} , \qquad F_{t x }{}{}^{(2)} = - A_x{}{}^{(1 )} A_t{}{}^{(3)} , 
\ee
we obtain the following relation binding the components of the ansatz if $A^{(1)}_x=w(z)\ne 0$
\be
\bigg( - \frac{(2 + \alpha)}{\alpha}~\xi(z) + \frac{2}{\alpha} S_D~(1-z^2) \bigg)~A_\phi (z)~g^{\phi \phi} -
\bigg( - \frac{(2 + \alpha)}{\alpha}~\eta(z) + \frac{2}{\alpha} \mu_D \bigg)~\varphi (z)~g^{tt} = 0.
\ee
This relation between $A_\phi(z)$ and $\phi(z)$ is valid for condensed state. On the other hand, for
$A^{(1)}_x=0$ equation (\ref{rel}) is identically fulfilled and both components of $A_\mu$ evolve
independently. 

\begin{figure}
\includegraphics[width=0.5\linewidth]{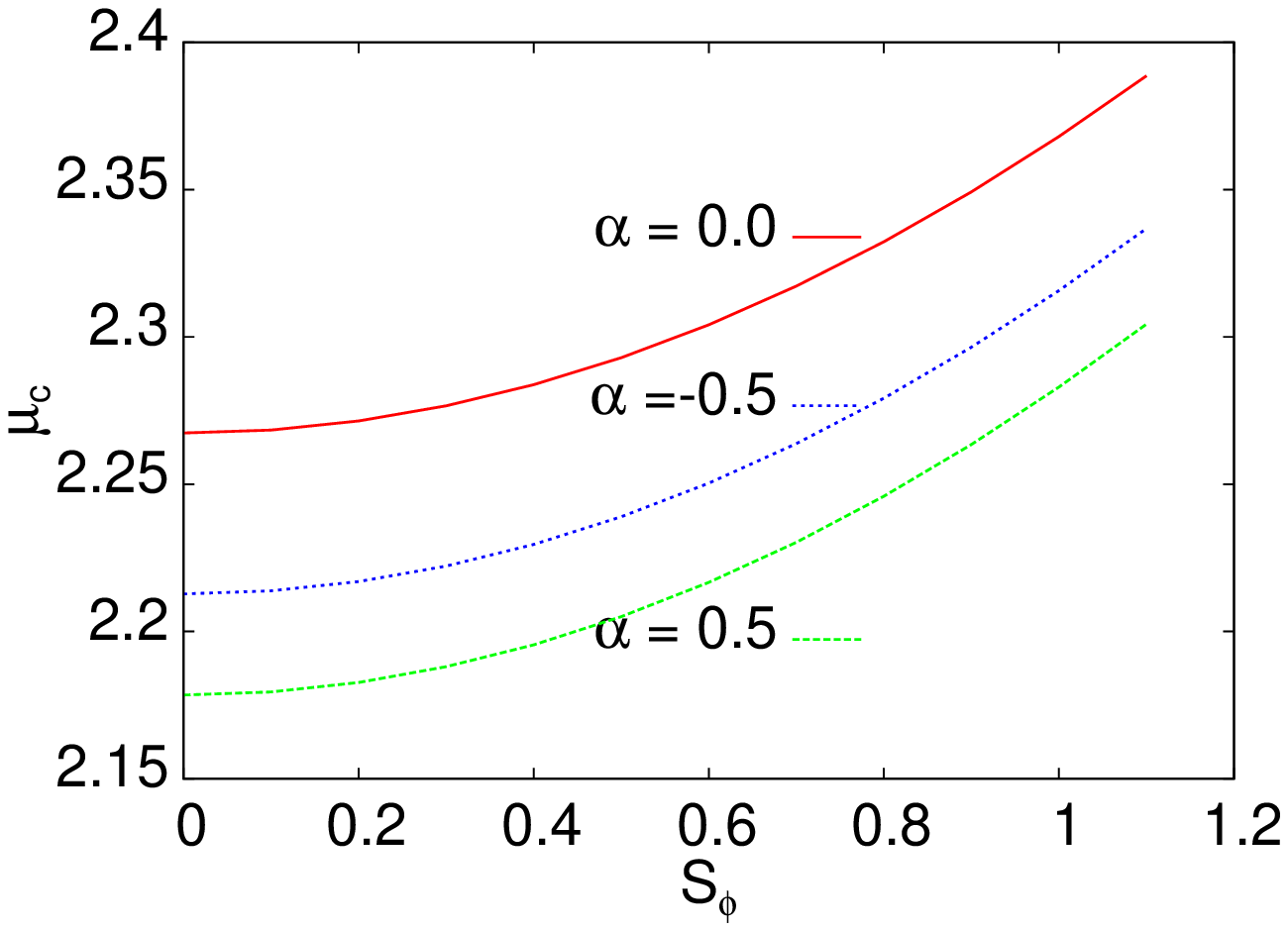}
\includegraphics[width=0.5\linewidth]{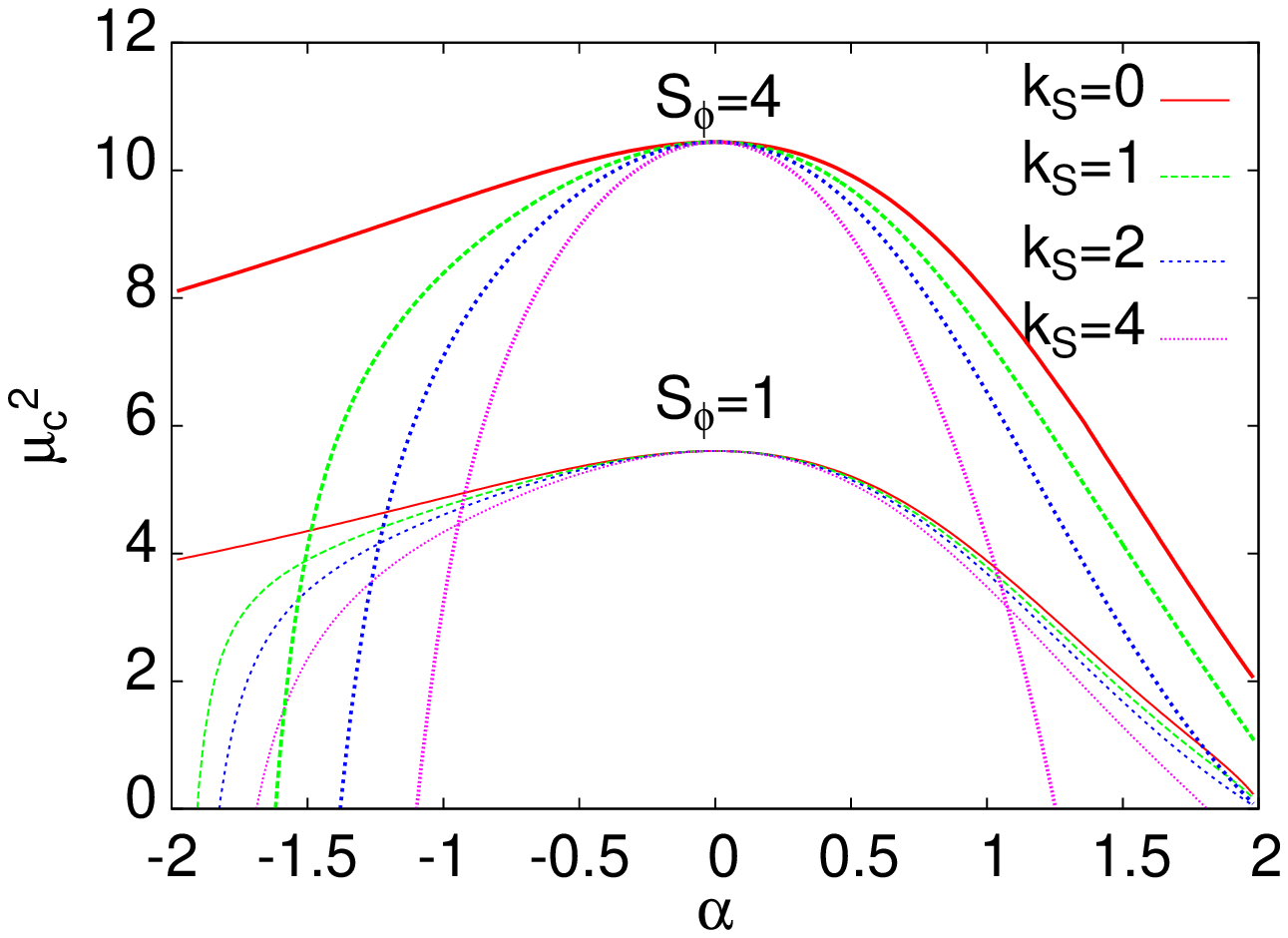}
\caption{In the left panel we illustrate the dependence of the critical value of $\mu_c$ 
 of the p-wave $SU(2)$ Yang-Mills superconductor on the velocity for a few values
of the coupling constant $\alpha$, for fixed $\mu_D=S_D=0$. The right panel shows the dependence of the critical 
chemical potential squared on the coupling 
$\alpha$ to the {\it dark matter}, for $k_\mu=0$, two values of velocity $S_\phi=1,~ 4$ and 
a number of parameters $k_S=S_D/S_\phi$.
 }
\label{fig3}
\end{figure}

\begin{figure}
\includegraphics[width=0.5\linewidth]{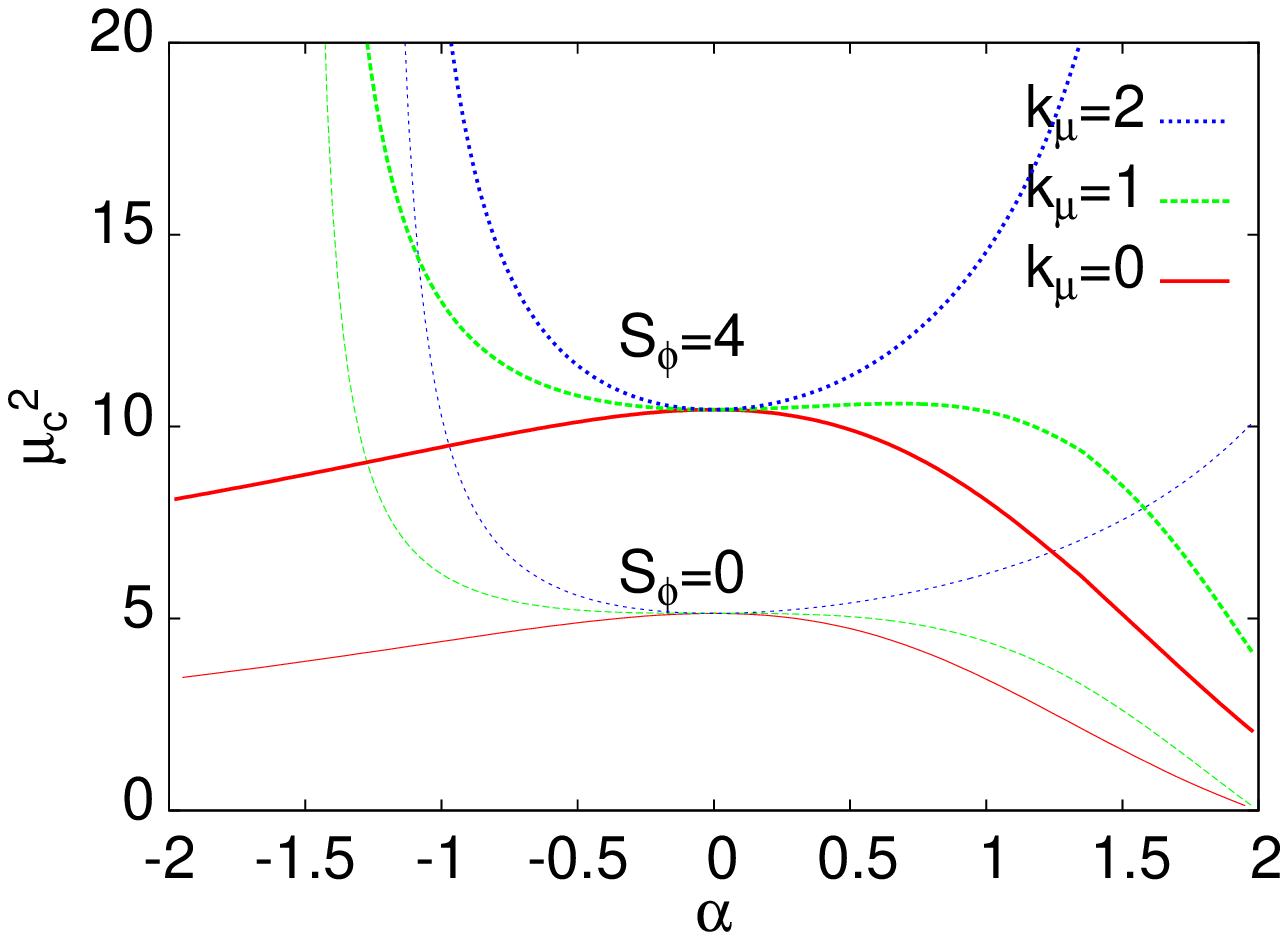}
\includegraphics[width=0.5\linewidth]{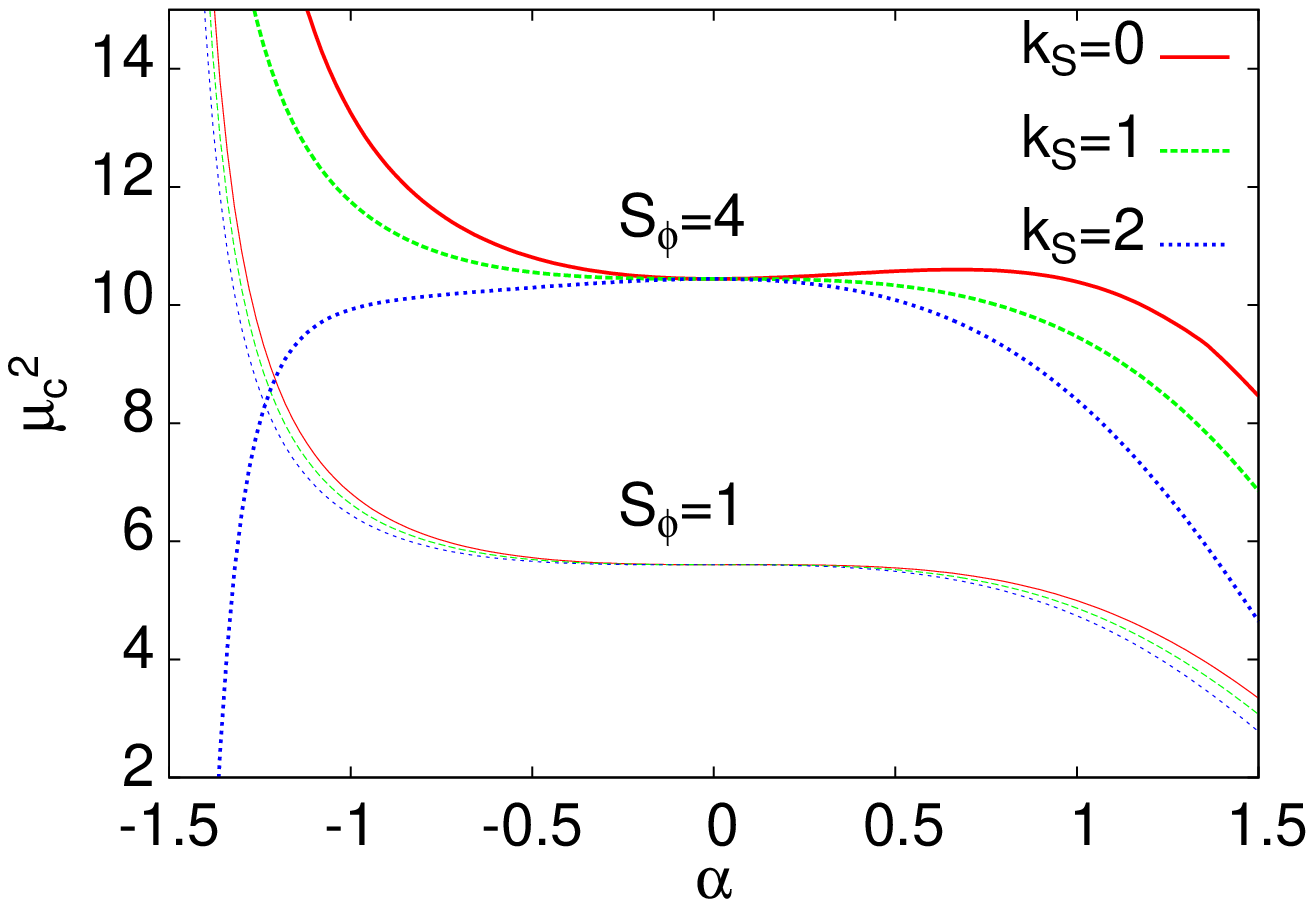}
\caption{ The dependence of the squared critical chemical potential on the $\alpha$-coupling 
to the {\it dark matter}, for $k_\phi=0$, a number of parameters $k_\mu=\mu_D/\mu_c$ and two 
values of superflow velocity $S_\phi=0,4$ (left panel). The right panel presents the similar dependence
for $S_\phi=1,4$ and a few values of  the velocity of dark matter 
parametrized by $k_S=S_D/S_\phi$ and for $k_\mu=1$.
Note that the effect of the {\it dark matter} velocity is relatively small for the small values of $S_\phi$ 
and dramatically increases
for the large values of the superflow.  }
\label{fig4}
\end{figure}

The above equations enable us to rewrite the equation (\ref{w1})  as
\ben
w''(z) &+& \bigg( \frac{f'(z)}{f(z)} + \frac{1}{z} \bigg)~w'(z) 
+ \mu^2~\frac{[\beta - (1-\tilde{\alpha})k_\mu]}{\tilde \alpha}~
\frac{1}{f(z)~z^2}~w(z)  \\ \nonumber
&-& S^2_\phi (1-z^2)^2~\frac{[\beta - 
(1-\talpha)~k_S}{\tilde \alpha} ~\frac{1}{f^2(z)~z^4}~w(z) = 0,
\een
where we have denoted by $k_\mu=\mu_D/\mu$ the ratio between chemical potentials of
dark and visible sector, while by $k_S=S_D/S_\phi$ similar ratio of the velocities,
$\beta=1+\alpha^3/8$.
The obtained equation for the field $w(z)$ is valid close to the critical value of the 
chemical potential $\mu_c$ and in fact constitutes an equation for its determination.
To find $\mu_c$, we 
correct the solution for $w(z)$ close to the boundary $z \rightarrow 0$
by  defining the trial function $G(z)$
\be
w(z) \sim <\cO>~z^2~G(z),
\ee
with $G(z)=1-az^2$ fulfilling the appropriate boundary conditions 
$G(0)=1$ and $G'(0)=0$ and $<\cO>=w_2$. 
Equation for $w(z)$ can be rewritten in the form adequate to study the Sturm-Liouville
eigenvalue problem \cite{sio10}
\be
(P(z)~G'(z))' - Q(z)~G(z) + {\Lambda}^2~R(z)~G(z) = 0,
\ee
where we have defined 
\ben
\Lambda^2 &=& \mu^2~\frac{\beta-(1-\tilde{\alpha})k_\mu}{\tilde \alpha}\\
P(z) &=& z^5~f(z),\\
Q(z) &=& -f(z)~\bigg( 4 z^3 + 2z^4 ~\frac{f'(z)}{f(z)} 
-S_\phi^2 (1-z^2)^2~\frac{[ \beta - (1-\talpha)~k_S]}{\talpha~f^2(z)} \bigg),\\
R(z) &=& z^3 .
\een
This equation allows us to find the  minimum eigenvalue of $\Lambda^2$, by the method 
of minimizing the following  functional with respect to $a$
\be
{\Lambda}^2 = 
\frac{\int_0^1 dz~[G'(z)^2~P(z) + Q(z)~G^2(z)]}{\int_0^1 dz~R(z)~G^2(z)}.
\ee
It has to be noted that the critical chemical potential $\mu_c$ we are looking at,
depends on the parameters $k_\mu$ and $k_S$ as well as on the velocity $S_\phi$ and 
the $\alpha$-coupling constant to {\it dark matter} sector. Under the adopted approximations 
the dependence
on $\mu_c(\alpha)$ has two sources: one is the direct dependence of $\Lambda(\alpha)$ and the other
is subject to the function $Q(z)$, in the considered functional.  The results are shown in the  figures \ref{fig3}
and \ref{fig4}.

In view of the dependence of the critical chemical potential of the p-wave
superfluid on four parameters ($\alpha$, $S_\phi$, $k_S$ and $k_\mu$), we shall only show 
some of the results.  
We start with the dependence of $\mu^2_c$ on the superfluid velocity $S_\phi$,
for a few values of the $\alpha$-coupling to the {\it dark matter} sector and for $k_S=k_\mu=0$. 
 The results are shown in the left panel of the figure \ref{fig3}. 

The right panel of the figure \ref{fig3} shows the dependence of $\mu^2_c$ on $\alpha$
for zero value of the dark matter chemical potential $\mu_D=0$, two values 
of the velocity  $S_\phi=1, 4$ and a few values of $k_S$
for each of the $S_\phi$.
For some values of $\alpha$, the Sturm-Liouville eigenvalue $\Lambda$ drops below zero, which
is unphysical. This means absence of the superconducting solution in these parameter ranges. Our equations
are valid close to critical value of chemical potential, but otherwise are exact. The lack of
the solution means that for those parameters, no matter how big will be the chemical potential, the
system will stay insulating. The increase of the {\it dark matter} velocity $S_D$ ($k_S=S_D/S_\phi$) 
generally
decreases the value of $\mu_c$ thus making the transition easier to appear, until one reaches 
zero value of $\mu_c$.
This conclusion is generally true also for non-zero value of the {\it dark  matter} chemical potential, 
as is visible in the right
panel of the figure \ref{fig4}, albeit the detailed dependence $\mu_c(\alpha)$ is different.  
The fact that $\mu_c$ increases with the superflow $S_\phi$ indicates the adverse effect 
of the latter parameter on the transition.

The left panel of figure \ref{fig4}  envisages the dependence of $\mu_c^2$ on the $\alpha$-coupling constant
of {\it dark matter} for $k_S=0$, a number of parameters $k_\mu=\mu_D/\mu_c$ and the two 
values of the superflow velocity $S_\phi=0,~4$. For a constant $\alpha\ne 0$, the growth of the {\it dark matter}
chemical potential $\mu_D$ generally increases $\mu_c$, thus making the condensation harder to appear. 
The right panel of figure \ref{fig4} shows the similar dependence of $\mu_c^2$ on $\alpha$ 
for $S_\phi=1,4$ and a few values of  the velocity of {\it dark matter} parametrized by $k_S=S_D/S_\phi$ but 
contrary to the figure  \ref{fig3} for $k_\mu=1$.
Note that the effect of the {\it dark matter} velocity is relatively small for small values of 
$S_\phi$ and dramatically increases
for the elevated superflow. In the latter case the detailed behavior strongly depends 
on the {\it dark matter} velocity, $i.e.$, on $k_S$. 

\subsubsection{Critical phenomena in p-wave superfluid model}
The equations of motion for $\varphi(z)$-component of the Maxwell field are the same 
as analyzed in \cite{rog16}, 
where holographic p-wave superconductor with {\it dark matter} sector have been
studied. We refer the readers to this reference for the particulars of calculations 
and figures describing 
the elaborated quantities. However in this section we present only a bird eye view 
on the problem in question.

When $\mu \rightarrow \mu_c$, the condensate operator value is small but finite 
and the equation for time component of Maxwell potential is given by
\be
\varphi''(z) + \bigg( \frac{f'}{f} + \frac{1}{z} \bigg)~\varphi'(z) - 
\frac{<\cO>^2~z^2~F^2(z)}{\tilde \alpha~f}~\varphi(z)=0.
\label{cr0}
\ee

Calculations analogous to those presented previously in ~\cite{rog16} lead to the following expression of the mean value of the condensation operator:
\be
<\cO> \simeq ~\Delta~(\mu - \mu_c)^\frac{1}{2},
\label{opsc}
\ee
where the pre-factor $\Delta=\sqrt{\frac{\tilde{\alpha}}{\mu_c\xi(0)}}$,~the exact value of $\xi(0)$ 
is given in \cite{rog16}.
The form of the equation (\ref{opsc}) envisages  the fact that the p-wave holographic
superfluid critical phenomenon represents the second order phase transition for which the critical 
exponent has the mean field value $1/2$. 

On the other hand, the charge density is found to linearly depend on $\mu$
\be
\rho = (\mu - \mu_c)~{\tilde{ D}},
\ee
where the quantity $\tilde{ D}$ is constant independent on $\alpha$.

\subsubsection{$A_\phi$ in $SU(2)$ p-wave holographic model}

In this subsection let us consider the velocity of p-wave superfluid current. The adequate equation 
of motion for the spatial component of Maxwell potential $A_\phi$ is
given by
\be
A_\phi'' (z) - \frac{1}{z}~A_\phi' (z) - \frac{w^2(z)}{\tilde \alpha~f(z)~z^2} ~A_\phi(z)= 0.
\ee
As in the preceding sections, we set again the ansatz for $w(z) = <\cO>~z^2~F(z)$ and 
reduce the relation to the form
\be
A_\phi'' (z) - \frac{1}{z}~A_\phi' (z) - \frac{<\cO>^2~z^2~F^2(z)}{\tilde \alpha~f(z)} ~A_\phi(z)= 0.
\ee
The function $F(z)$ obeys the standard boundary conditions $F(0)=1,~F'(0)=0$.
To proceed further, let us approximate $A_\phi$ near the critical point. It yields
\be
A_\phi \simeq S_\phi~(1-z^2) + <\cO>^2~\chi(z) + \dots
\ee
As in the previous case we expand $\chi(z)$, near the boundary of AdS space-time in a series provided by
\be
\chi(z) = \chi(0) + \chi'(0) z+ \frac{1}{2!}~\chi''(0) z^2 + \dots,
\ee
which leads to the relation of the form
\be
\chi''(z) - \frac{1}{z}~\chi'(z) = \frac{F^2(z)~z^2}{\tilde \alpha~f(z)}~(1-z^2)~S_\phi(z) 
+ \cO \bigg(<\cO>^{n\geq 2}\bigg).
\ee
Then, the above equation can be rewritten in the form
\be
(p(z)~\chi'(z))' + q(z)~F^2(z) =0,
\ee
where we have set
\be
p(z) = \frac{1}{z}, \qquad q(z) = - \frac{F^2(z)~z}{\tilde \alpha~f(z)}~(1-z^2)~S_\phi.
\ee
Integrating  the above relation and using the condition $\chi'(0) = 0$, one gains
\be
\chi''(0) = \frac{\chi'(z)}{z} \mid_{z \rightarrow 0} = 
-S_\phi \int_0^1 dz~\frac{(1-z^2)~z~F^2(z)}{\tilde \alpha~f(z)}.
\ee
Having in mind $z^2$-order terms, we can find that the superfluid current implies
\be
J_\phi = S_\phi + S_\phi~\frac{<\cO>^2}{\tilde \alpha}\int_0^1 dz~\frac{(1-z^2)~z~F^2(z)}{f(z)}~.
\label{curr-p-sol}
\ee
Consequently, the relation describing $A_\phi(z) $ yields
\be
A_\phi(z) = S_\phi~(1-z^2) - 
\frac{S_\phi~<\cO>^2}{\tilde \alpha}~z^2~\int_0^1 dx \frac{(1-x^2)~F^2(x)}{f(x)}.
\ee

 Taking into account the dependence of the pre-factor of $<\cO>$ (see the equation (\ref{opsc})) 
on $\talpha$, one finds that the {\it dark sector} does not directly effect the current of holographic
p-wave superconductor. Contrary to the s-wave case, the current depends on {\it dark matter}
indirectly {\it via}  $\mu_c=\mu_c(\alpha,\mu_D,S_D)$.

\section{Holographic s-wave superfluid in black hole background}
In this section we take up the problem of three dimensional s-wave holographic superfluid at a certain temperature. 
In order to scrutinize
the question one analyzes the background of five-dimensional AdS black hole given by the line element
\be
ds^2 = - f(r)~dt^2 + \frac{dr^2}{f(r)} + \frac{r^2}{L^2}~(dx^2 + dy^2 + dz^2),
\ee
where $f(r) = r^2/L^2 - r_+^4/r^2 L^2$. In what follows, without loss of generality we set $L=1$.
The Hawking temperature for the black hole is equal to $T_{BH} = r_+/\pi$ and defines the temperature
$T$ at the boundary.  
We assume that the non-zero components of the Maxwell fields are given by
$A_t(r) = \phi(r),~A_y(r)$. The equations of motion read 
\ben
\psi''(r) &+& \bigg( \frac{3}{r} + \frac{f'}{f} \bigg)~\psi'(r) - \bigg(  \frac{m^2}{f(r)}  - \frac{q^2~\phi^2(r)}{f^2(r)}
 + \frac{q^2~A^2_y(r)}{r^2~f(r)} \bigg)~\psi(r) = 0,\\
\phi''(r) &+& \frac{3}{r}~\phi'(r) - 2 ~\frac{q^2~\psi^2(r)}{\talpha ~f(r)} ~\phi(r)= 0,\\
A''_y(r) &+& \bigg( \frac{1}{r} + \frac{f'}{f} \bigg)~A'_y(r) - 2 ~\frac{q^2~\psi^2(r)}{\talpha~r^3~f(r)}~A_y(r)
 = 0,
\een
where the prime denotes the derivative with respect to $r$-coordinate. We again note that the dependence
between $\phi(r)$ and $A(r)_\mu$ is induced by the condensation of the $\psi(r)$ field.

The close inspection of the above relations envisages that  in $z= r_+ /r$-coordinates, 
they are given by
\ben
\psi''(z) &+& \bigg( \frac{f'}{f} - \frac{1}{z} \bigg)~\psi'(z) + 
\frac{q^2~\phi^2(z)}{r_+^2z^4~f^2(z)}~\psi(z) - \frac{q^2~A^2_y(z)}{~r_+^2 z^2~f(z)}~\psi(z) -
 \frac{m^2}{z^4~f(z)}~\psi(z) = 0,\\ \label{eqphi}
\phi''(z) &-& \frac{1}{z}~\phi'(z) - 
2 ~\frac{q^2~\psi^2(z)}{\talpha ~f(z)~z^4}~\phi(z) = 0,\\
A''_y(z) &+& \bigg( \frac{1}{z} + \frac{f'}{f} \bigg)A'_y(z) - 
2 ~\frac{q^2~\psi^2(z)}{\talpha~r_+^3~f(z)~z}~A_y(z) = 0,
\label{eqAyz}
\een
where now the prime is bounded with taking derivative with respect 
to $z$-coordinate. We have also denoted by $f(z)$ the relation
\be
f(z) = ~\bigg( \frac{1-z^4}{z^2}\bigg).
\ee

\subsection{Critical temperature for s-wave superfluids}

For $T \rightarrow T_c$ the condensate is very small $\psi(z)\rightarrow 0$. The value 
of the horizon radius for the black hole with temperature $T_c$ is denoted by $r_{+c}$. 
The asymptotic boundary conditions, as $r$ tends to infinity, are the same as given by 
the equation (\ref{bc}), with the replacement of $A_\phi$ and $J_\phi$ by 
$A_y$ and $J_y$, respectively.

The equation (\ref{eqphi}) for the $\phi$ field near the critical point reduces to the relation
\be
\phi^{''}(z) - \frac{\phi'(z)}{z} \simeq 0,
\ee
which has the general solution of the form $\phi(z)=c_1+c_2z^2$. The boundary conditions 
imposed on Maxwell t-component gauge field $\phi(1)=0$,
enable us to find that 
$$
\phi \simeq \rho r_{+c}^{-2}(1 - z^2).$$

Consequently, the inspection of the spatial component of the Maxwell gauge field, 
near the critical temperature implies
\be
A''_y(z) + \bigg( \frac{f'}{f} + \frac{1}{z} \bigg)~A'_y(z) \simeq 0,
\ee
with the condition $A_y(1)=0$, we get $A_y = S_y$. We are looking for the  function $\psi(z)$ near
 the boundary $z \rightarrow 0$ of the considered spacetime. It is approximated 
by the expression
\be
\psi(z) \mid_{z \rightarrow 0} = \frac{<\cC>}{r^\Delta_+}~z^\Delta~F(z),
\label{apppsi}
\ee
where we have to set $F(0) =1,~F'(0) =0$.  Finally, near the critical temperature one arrives at
\ben
F''(z) &+& F'(z)~\bigg[ \frac{2 \Delta}{z} + \bigg( \frac{f'}{f} - \frac{1}{z} \bigg) \bigg]
 + \la^2~\frac{(1-z^2)^2}{f^2(z)~z^4}~F(z) +\\ \nonumber
&+& F(z)~\bigg[
\frac{\Delta~(\Delta -1)}{z^2} + \frac{\Delta}{z}~\bigg( \frac{f'}{f} - \frac{1}{z} \bigg)
 - \frac{m^2}{z^4~f(z)} - \frac{q^2~S^2_y}{~r^2_+z^2~f(z)} \bigg] = 0,
\een
with the parameter $\lambda^2=\rho^2~q^2/~r^6_+$. 
Repeating the procedure leading to the Sturm-Liouville functional, we get
\be
\la^2 = \frac{\int_0^1 dz~[F'(z)^2~p(z) + q(z)~F^2(z)]}{\int_0^1 dz~r(z)~F^2(z)},
\label{sl-4}
\ee
where the introduced functions are defined as  
\ben
p(z) &=& z^{2 \Delta-1}~f(z),\\
q(z) &=& - z^{2 \Delta-1}~f(z)~\bigg[
\frac{\Delta~(\Delta -1)}{z^2} + \frac{\Delta}{z}~\bigg( \frac{f'}{f} - \frac{1}{z} \bigg) 
- \frac{m^2}{z^4~f(z)} - \frac{q^2~S^2_y}{~r^2_+~z^2~f(z)} \bigg] ,\\
r(z) &=& \frac{z^{2\Delta-5}}{f(z)}~q^2~(1-z^2)^2.
\een
Interestingly,  the relation for $\la^2$ which serves as a condition for the critical temperature
of the superconductor itself depends parametrically on the temperature of the considered 
black hole. This is a direct consequence of the fact that the Hawking black hole temperature 
$T_{BH} = r_+/\pi$ and $\la_{min} = \rho q/r^3_+$ enter the function $q(z)$ above. 
In order to find the transition temperature $T_c$, one ought to elaborate the self-consistent 
solution of the transcendental equation (\ref{sl-4}), as the  
 critical temperature is related to  $\la_{min}$  $via$ relation
\be
T_c = (\rho)^{\frac{1}{3}}~\bigg( \frac{1}{\pi^3~\la_{min} }\bigg)^{\frac{1}{3}}.
\label{tc-la-min}
\ee
with $\la_{min}$ depending  on $T_c$.
This behavior is characteristic to the superconductor carrying the current or
superfluid with non-zero velocity of the condensate. Without superflow, $S_y=0$,
one gets simple equation for $T_c$.

\begin{figure}
\includegraphics[width=0.5\linewidth]{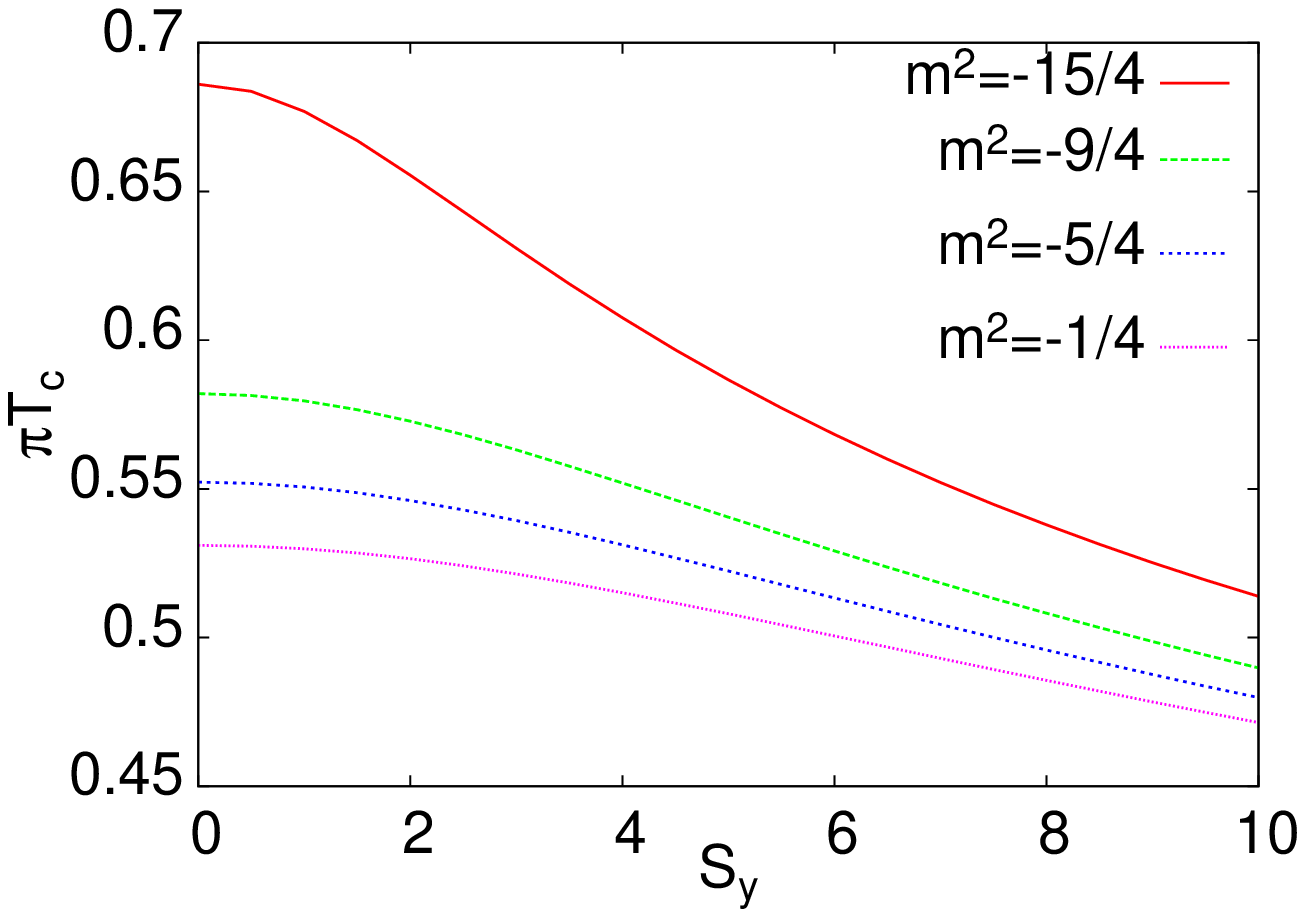}
\includegraphics[width=0.5\linewidth]{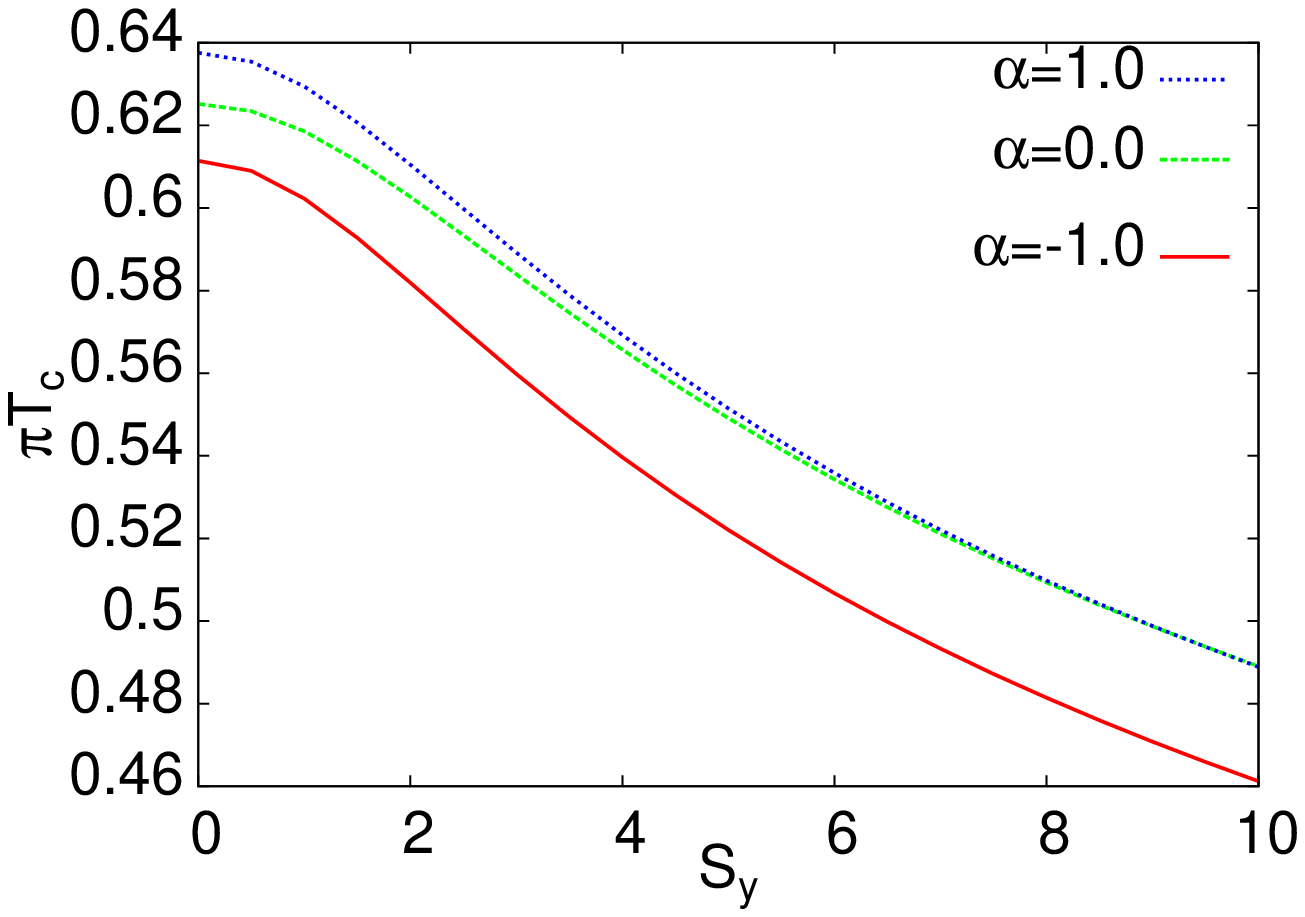}
\caption{ The dependence of the self-consistently calculated transition temperature
of the current carrying superconductors of s-wave symmetry, equation (\ref{tc-la-min}) (left panel) 
and the p-wave symmetry, equation (\ref{tc-pw-su2}) (right panel), on  the superflow 
velocity $S_y$ for a number of mass parameters  $m^2$
or couplings $\alpha$, respectively. For numerical  purposes
we have taken $q=1$ and $\rho=1$. 
In the right panel we have set $k_S=0$ and $k_\rho=0$.}
\label{fig5}
\end{figure}

We have solved the resulting equation numerically in a self-consistent way.
The results of the dependence of $T_c$ on $S_y$ for a number of $m^2$ values 
are presented in figure \ref{fig5}. The increase of the velocity results in the rise
of the eigenvalue of the Sturm-Liouville problem and thus by the equation  (\ref{tc-la-min})
to decrease of the superconducting transition temperature. Quantitatively it agrees with
the known phenomenology of real superconductors. 

\subsection{Condensation operator values}
Here we study  the influence of {\it dark matter} 
sector on the condensation operator for s-wave holographic superfluid.
The equation of motion for $\phi$ is independent of the spatial component of the 
Maxwell field potential $A_y$ and has the form studied in \cite{nak15a}.
So we address here only the main features of the derivation, referring the reader 
the the previous work, for details.

The equation of motion for the time component of the Maxwell field, near  the critical 
temperature is obtained from by introducing (\ref{apppsi}) into (\ref{eqphi}) and is given by
\be
\phi''(z) - {\phi'(z) \over z} = {2~q^2~r_+^2~\phi(z) \over \tilde{\alpha}~f(z)}~z^{2\Delta -4}
~\frac{<\cC>^2}{ r_+^{2\Delta}}~F^2(z).
\ee
Proceeding as in \cite{nak15a} for $T \rightarrow T_c$, the condensation operator
 $<\cC>$ is found as 
\be
<\cC> = \sqrt{2~\tilde{\alpha} \over \cB}~(\pi~T_c)^\Delta~\sqrt{1 - {T \over T_c}},
\ee
where
$\cB$ is given~\cite{nak15a} by
\be
\cB = 2\int_0^1 dz~{q^2~r_+^2 \over f(z)}~(1-z^2)~z^{2\Delta-5}~F^2(z).
\ee
The mean value of the condensation operator depends on the $\alpha$-coupling constant of the
{\it dark matter} sector. The bigger $\alpha$-coupling one takes into account, the smaller is the
value of  the condensate mean value operator  $<\cC>$. In addition operator $<\cC>$ depends on $\alpha$
{\it via} $T_c(\alpha)$.

\subsection{$A_y$ in s-wave superfluid at given temperature}
Having in mind the relation (\ref{apppsi}) the equation of motion for $A_y(z)$ (\ref{eqAyz}) 
near $T \rightarrow T_c$ reads
\be
A''_y(z) + A'_y(z)~\bigg( \frac{f'}{f} + \frac{1}{z} \bigg) - 
\frac{ 2~q^2~A_y}{\talpha}<\cC>^2~\frac{z^{2\Delta-1}~F^2(z)}{f(z)~r_+^{2\Delta+1}} \simeq 0.
\ee
To proceed further, we assume that
\be
A_y(z) \simeq S_y + <\cC>\chi(z),
\ee
and restrict our consideration to the terms of order $<\cC>$. Consequently  we obtain
\be
\chi''(z) + \chi'(z)~\bigg( \frac{f'}{f} + \frac{1}{z} \bigg) -
 2 \frac{ q^2~<\cC>~S_y}{\talpha~f(z)~r_+^{2 \Delta+3}}~z^{2 \Delta -1}~F^2(z) \simeq 0.
\ee
Like in the preceding sections,  expansion of $\chi(z)$ in series near $z=0$ 
and the comparison of terms in the expansion of $z^2$-order, enable us to write
the relation binding $A_y(z)$ and the super-current.  It implies
\be
A_y(z) = S_y - z^2~\frac{<\cC>^2~S_y~q^2}{\talpha~r_+^{2 \Delta +3}}~\int_0^1 dx~x^{2\Delta }~F^2(x).
\ee
One can point out that, as in the AdS solitonic background the {\it dark matter} sector 
increases the value of the super-current $J_y(z)$, while its dependence on $S_y$ remains nearly
linear in a close analogy to the s-wave superconductor in a solitonic background.
On the other hand, the current does depend on the coupling constant to {\it dark matter} sector only through $T_c(\alpha)$, 
as is visible from the dependence of $<\cC>^2 \propto \talpha$. 
More importantly, we have found the dependence of the current on temperature
in the linear form $J_y \propto (T_c-T)$, sometimes called the Onnes relation. 
This fact is in a significant contrast to the Ginzburg-Landau analysis
of $J_c$ for thin superconducting films, i.e., effectively two-dimensional systems, were one has $J\propto (T_c-T)^{3/2}$.
The experimental relevance of these results will be discussed later.

\section{Holographic p-wave superfluid in black hole background}
In this section we shall pay attention to the holographic p-wave superfluid case at finite temperatures. 
As in the previous sections, one can consider at least two models of p-wave holographic
 superfluids, i.e., the Maxwell vector and $SU(2)$ one.
The same arguments as quoted in the section concerning holographic p-wave superfluid models 
in the AdS solitonic background, can be implemented in this case.
Namely, the Maxwell vector model reduces to the s-wave case when one elaborates 
real vector field $\rho_x$. Therefore we restrict our considerations to the $SU(2)$ model.
Taking into account five-dimensional AdS black hole background we choose gauge fields 
components as
\ben  \label{ch1}
A &=& \phi(r)~\tau^3~dt + A_{y}(r)~\tau^3~dy +w(r)~\tau^1~dx, \\ 
B &=& \eta(r)~\tau^3~dt  + \xi(r)~\tau^3~dy.
\een
The $z$-dependent  equations of motion $x(1)$,~$y(3)$ and $t(3)$ can be written as           
\ben \label{w1a}
w''(z) &+& \bigg( \frac{f'(z)}{f(z)} + \frac{1}{z} \bigg)~w'(z) 
+ \frac{\phi (z)[\phi(z)-\frac{\alpha^2}{4}~\eta(z)]}{\tilde \alpha~
f(z)~z^2}~w(z) + \\ \nonumber 
&-& \frac{A_y (z)[A_y(z)-\frac{\alpha^2}{4}~\xi(z)]}{r_+^4~\tilde \alpha~
f^2(z)~z^4}~w(z) = 0,\\ \label{w2b}
A_y''(z) &-& \frac{1}{z}~A_y'(z) - \frac{w^2(z)}{r_+^2~\tilde \alpha~z^2~f(z)}~A_y(z)
= 0,\\ \label{php}
\phi''(z) &-& \frac{1}{z} ~\phi'(z) - \frac{w^2(z)}{r_+^2~\tilde \alpha~z^2~f(z)}~\phi(z)
= 0,
\een
where the prime denotes the derivative with respect to $z$-coordinate, $f(z)$ is
described in the preceding section. The structure of the above equations is analogous to those
studied earlier, with obvious changes related to the SU(2) character of the gauge field components.
In particular their independence is lost due to the hairy structure, $w(z)\ne 0$,  of the black hole.
For $w(z)=0$, the equations (\ref{w2b}) and (\ref{php}) are independent and each of the components of the $A_\mu$ field
evolves independently, albeit in the same way as it is visible from the above equations of motion.

The choice of the gauge field components (\ref{ch1}) is the only consistent choice 
enabling the analytic considerations of the problem in question.
The set of the differential equations (\ref{w1a})-(\ref{php}) ought to be 
accomplished by the addition of the boundary conditions. One assumes that
on the black hole event horizon $\phi(1)=0$ and the condensation field is 
of a finite norm, which in turn requires that $w(r_+)$ should be finite.
By virtue of the above, we assume that the following will hold in $r$ space 
\ben
w &=& w_0 + w_1~(r-r_0) + w_2~(r-r_+)^2 + \dots,\\
A_y &=& A_{y (0)} + A_{y (1)} (r-r_+)^2 + \dots,\\
\phi &=&  \phi_{(1)}~(r-r_+) + \phi_{(2)}~(r-r_+)^2 + \dots,\\
\eta &=& \eta_{(0)} + \eta_{(1)} (r-r_+) + \eta_{(2)}(r-r_+)^2 + \dots,\\
\xi &=& \xi_{(0)} + \xi_{(1)} (r-r_+) + \xi_{(2)}(r-r_+)^2   + \dots,
\een
where $w_j$~$A_{y (j)},~\phi_{(j)},~\eta_{(j)},~\xi_{(j)}$, for $j = 0,~1,~2, \dots$ are  
constants. The Neumann-like
boundary condition are required to obtain every physical quantity finite.
When $r$ tends to infinity (one is close the boundary of the AdS space-time)
the fields behave as 
\be
\phi(r) \rightarrow \mu - \frac{\rho}{r^2}, \qquad w(r) \rightarrow w^{(0)} 
+ \frac{w^{(2)}}{r^2}, \qquad
A_y (r) \rightarrow S_y - \frac{J_y}{r^2},
\ee
From equations
\ben
\na_\mu B^{\mu t (3)} &+& \frac{\alpha}{2}~\na_\mu F^{\mu t (3)} = 0,\\
\na_\mu B^{\mu y (3)} &+& \frac{\alpha}{2}~\na_\mu F^{\mu y (3)} = 0,
\een
the dependence of $\phi (z)$ and $A_y(z)$ on the component of the {\it dark matter} 
sector $\eta(z)$ and $\xi(z)$ can be established.
Namely, having in mind the boundary conditions $\eta(1) = 0$ and $\phi(1) = 0$
we can identify the integration constants with {\it dark flow} and {\it dark density}. 
It results in the following
relations binding the ordinary and {\it dark matter} characteristics for the 
holographic p-wave superfluid
\ben \label{ch2}
\xi(z) &+& \frac{\alpha}{2} ~A_y= S_D,\\
\eta(z) &+& \frac{\alpha}{2}~\phi(z) = \frac{\rho_D}{r^2_+}~(1-z^2).
\een

For $z$ close to boundary, $z\rightarrow 0$,  we 
use the fact that $\phi(z)=(\rho/r^2_+)(1-z^2)$ 
and $A_y(z) =S_y$ and rewrite the equation (\ref{w1a}), having in mind the 
relations (\ref{ch2}). Consequently it yields
\ben
w''(z) &+& w'(z) \bigg( \frac{1}{z} + \frac{f'}{f} \bigg)  
+ \frac{1}{\talpha~z^4~f^2(z)} \bigg(\frac{\rho}{r^3_+}\bigg)^2 (1-z^2)^2~\bigg[
\beta - (1-\talpha)~k_\rho \bigg] w(z) \\ \nonumber
&-& \frac{~S_y^2}{r^2_+\talpha~z^2~f(z)}~\bigg[
\beta - (1-\talpha)~k_S \bigg] w(z) = 0,
\een
where we have defined
\be
k_S = \frac{S_D}{S_y}, \qquad k_\rho = \frac{\rho_D}{\rho}.
\ee
Inserting $w(z) \sim <\cR>~z^2~G(z)$, we can rewrite the above equation 
in the form characteristic for studies of the Sturm-Liouville variational 
problem, which results in
\ben
G''(z) &+& \bigg( \frac{5}{z} + \frac{f'}{f} \bigg)G'(z) +
\bigg[ \frac{4}{z} +\frac{2}{z} \frac{f'}{f} - \frac{~S_y^2}{r_+^2~\talpha~f(z)~z^2}~\bigg(
\beta - (1-\talpha)~k_S \bigg) \bigg] G(z) \\ \nonumber
&+& \Lambda^2~\frac{(1-z^2)^2}{f^2(z)~z^4}~G(z) = 0,
\een
where $\Lambda^2$ is provided by
\be
\Lambda^2 = \bigg( \frac{\rho}{r^3_+} \bigg)^2~[ \beta - (1 -\talpha)~k_\rho].
\ee
The Sturm-Liouville variational problem enables us to achieve the minimum 
eigenvalue of $\Lambda^2$ from the functional
\be
\Lambda^2 = \frac{\int_0^1 dz~[G'(z)^2~p(z) + q(z)~G^2(z)]}{\int_0^1 dz~r(z)~G(z)},
\ee
where we have defined the following quantities 
\ben
p(z) &=& z^5~f(z),\\
q(z) &=& - z^5~f(z)~\bigg[ \frac{4}{z^2} + \frac{2}{z}\frac{f'}{f}
 - \frac{~S^2_y}{r_+^2\talpha~z^2~f(z)}~\bigg[ \beta - (1-\talpha)~k_S \bigg] ,\\
r(z) &=& \frac{z~(1-z^2)^2}{f(z)}.
\een
The transition temperature is calculated as 
\be
\pi T_c= \left(\frac{\rho^2[\beta - (1 -\talpha)~k_\rho]}{\Lambda^2 }\right)^{1/6}.
\label{tc-pw-su2}
\ee 
In a close analogy to the s-wave superconductors discussed earlier, $T_c$ of the
SU(2) p-wave superconductor has also to be determined in a self-consistent way, as  
the function $q(z)$ above depends on $r_+$ and  makes $\Lambda$
to depend on $T_c$. 
\begin{figure}
\includegraphics[width=0.5\linewidth]{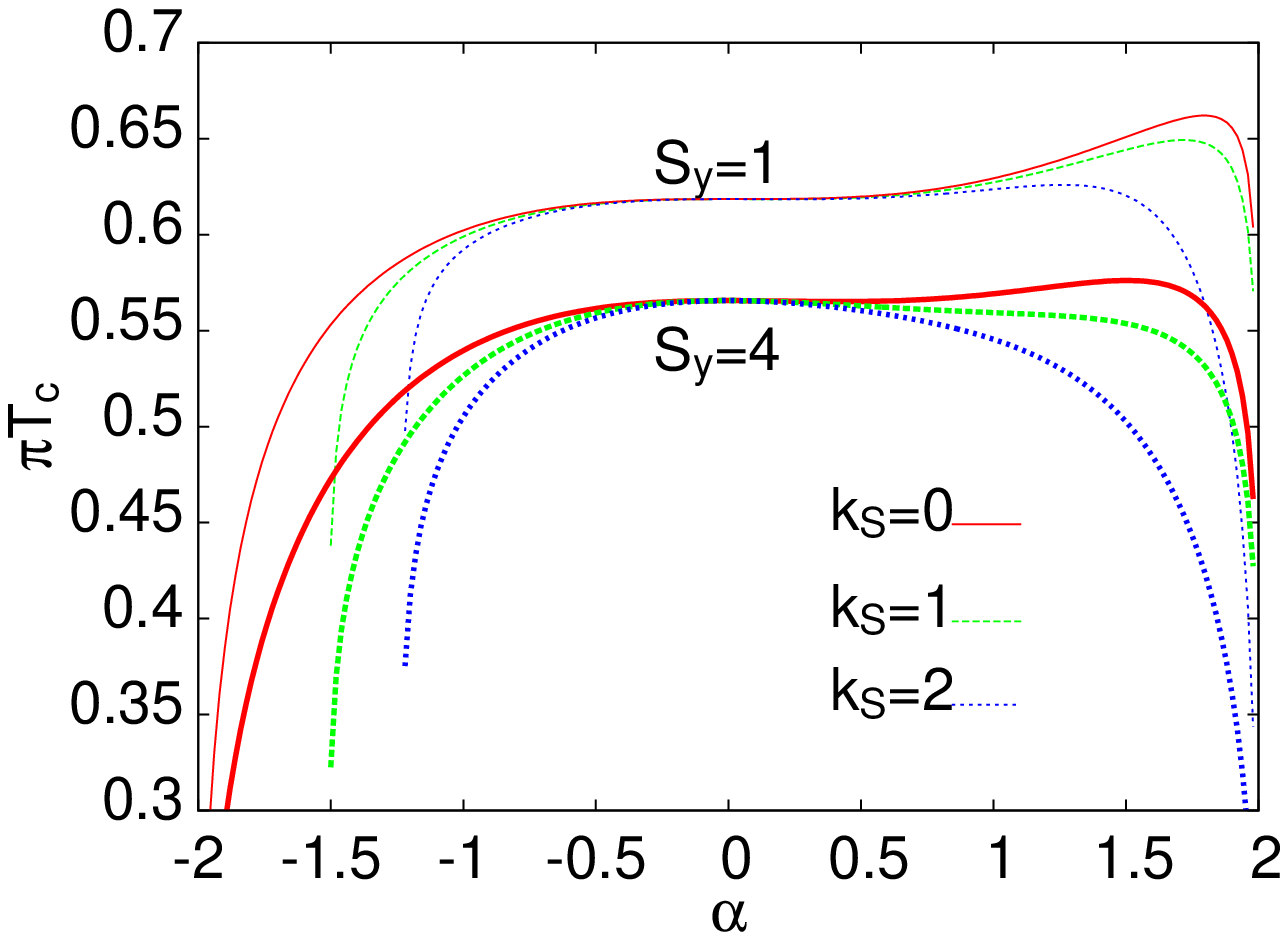}
\includegraphics[width=0.5\linewidth]{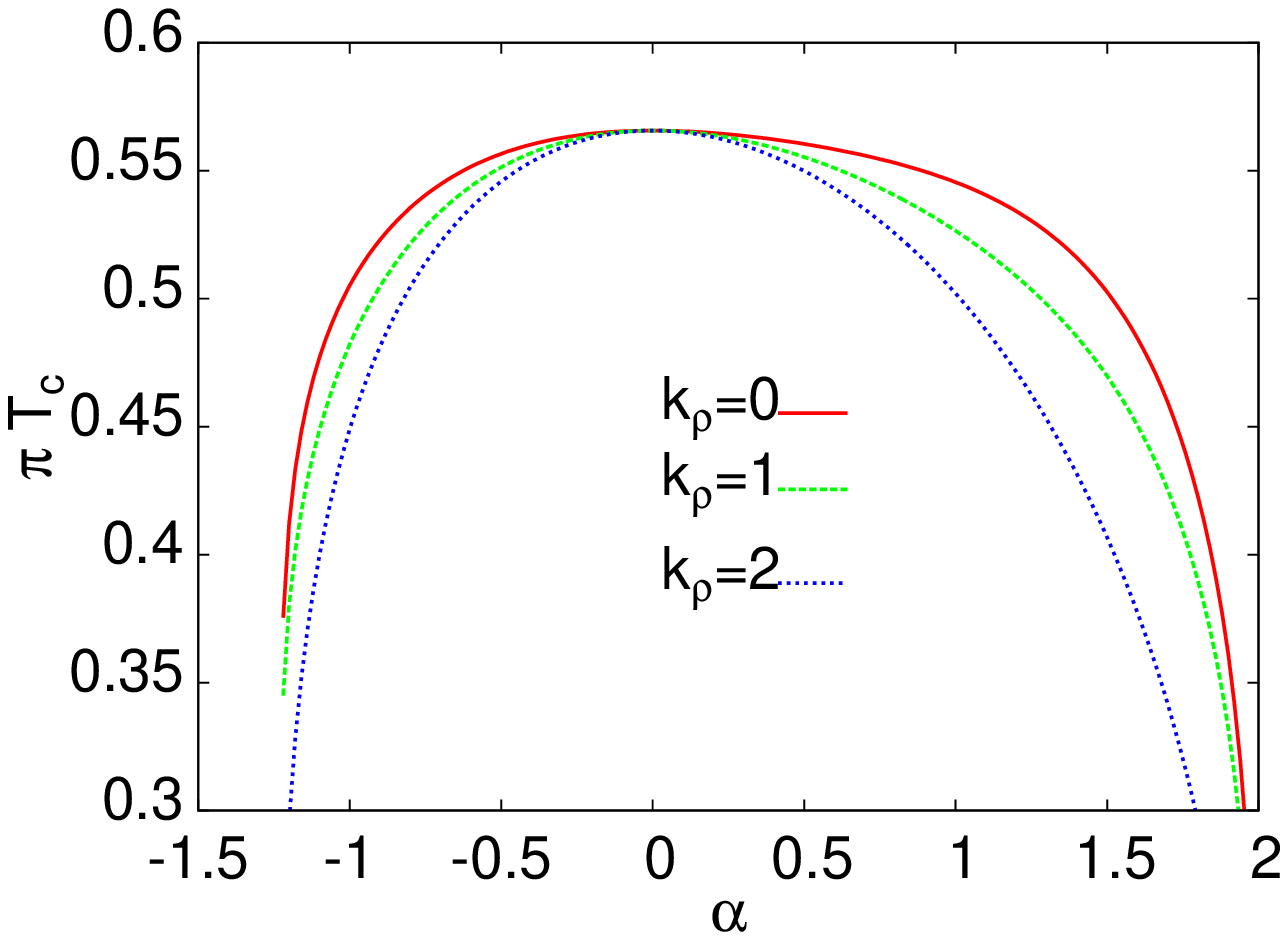}
\caption{In the left panel we depicted the dependence of the self-consistently calculated transition temperature
of the current carrying superconductor, equation (\ref{tc-pw-su2}),  on  the $\alpha$-coupling constant,
for two values of the velocity $S_y=1,~4$,~$k_\rho=0$ and a few values of the {\it dark matter} 
velocity $S_D$ parametrized here by $k_S=S_D/S_y$ is shown in the left panel.
The right panel shows similar dependence of $T_c(\alpha)$ for $S_y=4$,~ $k_S=2$
and three values of $k_\rho=0,1,2$. 
}
\label{fig6}
\end{figure}

The results of the self-consistent calculations of the $\alpha$ dependence
of the transition temperature ($\pi T_c$) are shown in the left panel of figure \ref{fig5} and in 
figure (\ref{fig6}). For 
the current carrying superconductor $T_c$ is calculated form the relation (\ref{tc-pw-su2}). 
We show the results for $k_\rho=0$, two values of the superfluid velocity $S_y=1,~4$ and 
three values of the dark matter  velocity $S_D$ parametrized here by $k_S=S_D/S_y$.
With increasing dark matter velocity $T_c$ diminishes and eventually it ceases
to exist (one gets negative solution for $\Lambda^2$) for negative values of $\alpha$.
The behavior is qualitatively the same for both values of $S_y$. 
The right panel of figure \ref{fig6} shows the similar dependence of $T_c(\alpha)$ 
for $S_y=4$, ~$k_S=2$ and three values of $k_\rho=\rho_D/\rho =0,1,2$. The increase 
of the {\it dark matter} density generally leads to the relatively small decrease of $T_c$, 
for a given coupling $\alpha<0$ and larger decrease, for $\alpha$ closer to the value $2.$

\subsection{Condensation value}
In this subsection we discuss the behavior of the condensation 
operator due to the presence of {\it dark matter} sector and its dependence 
on temperature, near the critical one, $T_c$. The form of the equation (\ref{php}) 
is the same as studied in \cite{rog16}, so for the details we refer  
to this work. 
For completeness of our discussion, we quote only the final result for the condensation operator $<\cR>$
\be
<\cR> = \sqrt{\frac{2\tilde{\alpha}\pi^2}{E}} T_c(\alpha)\sqrt{1+\frac{T}{T_c(\alpha)}}
~\sqrt{1 - \frac{T}{T_c(\alpha)}},
\ee 
where $E$ is given by the integral
\be
E = \int_0^1 dz~\frac{z^3~F^2(z)}{(1+z^2)}.
\ee
One concludes that $<\cR>$ depends on $\alpha$-coupling constant of the {\it dark matter} sector
directly {\it via}  factor $\tilde{\alpha}^{1/2}$ and indirectly through $T_c(\alpha)$. 
It also depends on the
density $\rho_D$ of the dark matter {\it via} $T_c$. 
The mean value of the operator $<\cR>$ can be interpreted as responsible for
 the pairing mechanism. The smaller vacuum
expectation value it has, the harder condensation happens. So we 
conclude that {\it dark matter} sector 
destructively influences  the condensation phenomena in p-wave 
superconductors for $\rho_D/\rho>1$. One has to notice, that in the presence
of the velocity the dependence of $<\cR>$ on $\alpha$ differs from that without
super-flow and the resulting $\alpha$ dependence of $T_c$ will be different
from that reported earlier~\cite{rog16}.

\subsection{$A_y$ in p-wave black hole holographic superfluid}
As in the preceding sections we want to find the velocity and the current of the y-directed superflow.
The equation of motion for $A_y(z)$  is of the form
\be
A_y'' (z)  + \bigg( \frac{1}{z} +\frac{f'}{f} \bigg)~A_y' (z) 
- \frac{w^2(z)~A_y(z)}{r_+^2~\talpha~f(z)~z^2} = 0.
\ee
For $T$ close to $T_c$ we correct $w(z)$ using an ansatz  $w(z) = <\cR>~z^2~F(z)$ 
and reduce the relation to the form
\be
A_y'' (z) + \bigg( \frac{1}{z} +\frac{f'}{f} \bigg)~A_y' (z)
 - \frac{<\cR>^2~z^2~F^2(z)}{r_+^2~\tilde \alpha~f(z)}~A_y(z) = 0.
\ee
To proceed, let us approximate $A_y(z)$ near the critical point by 
\be
A_y \simeq S_y + <\cR>~\chi(z) + \dots
\ee
As in the previous cases we expand $\chi(z)$, near the boundary of AdS
 space-time in a series provided by
\be
\chi(z) = \chi(0) + \chi'(0) z+ \frac{1}{2!}~\chi''(0) z^2 + \dots,
\ee
which leads to the relation 
\be
\chi''(z) + \bigg( \frac{1}{z} + \frac{f'}{f} \bigg)
~\chi'(z) = \frac{F^2(z)~z^2}{\tilde \alpha~f(z)}~(1-z^2)~S_y(z) + \cO \bigg(<\cR>^{n\geq 2}\bigg).
\ee
The above equation can be rewritten in the form
\be
(p(z)~\chi'(z))' + q(z)~F^2(z) =0,
\ee
where we set for $p(z)$ and $q(z)$ the following relations:
\be
p(z) = f(z)~z, \qquad q(z) = - \frac{<\cR>~z^3}{r_+^2~\tilde \alpha}~S_y.
\ee
Integrating  the above equation and using the condition $f(0) = 0$, enable us to find that
\be
\chi''(0) = \frac{\chi'(0)}{z} \mid_{z \rightarrow 0} = 
-\frac{S_y}{r_+^2~} \int_0^1 dz~\frac{z^3~F^2(z)}{\tilde \alpha}.
\ee
Thus we get
\be
A_y(z) = S_y - \frac{S_y~<\cR>^2 }{2r_+^2~\talpha}~z^2~\int_0^1 dx ~x^3~F^2(x),
\ee
where the term multiplying $z^2$ is interpreted as the superfluid current.
In analogy to the previously studied cases, the coupling constant to the {\it dark matter} sector 
cancels out and the current depends  on  $\alpha$ via transition temperature $T_c(\alpha)$
enters through $r_+$. Again, we find the linear in temperature, $(T_c-T)$  dependence, of the current.
The disappearance of the current at $T=T_c$ is a well established behavior. It is encouraging that the holographic
approach recovers this behavior.


\section{Summary and discussion}
\label{sec:conclusions}
In the paper we have considered the properties of holographic superfluid with the superflow of
the condensate, i.e., the situation when on the gravity side  one
accepts not only  t-component of the gauge fields but also takes into account the spatial one,  
as well.  The problem in question has been 
studied analytically by means of the Sturm-Liouville variational method. 
We have analyzed the s-wave and p-wave current carrying superfluids 
with {\it dark matter}  sector which 
has been described  by additional $U(1)$-gauge field coupled to the ordinary one, 
in the background of the AdS soliton (T=0) or black hole (T$\ne$ 0). We have elaborated the probe limit case.
According to the AdS/CFT dictionary the asymptotic ($r\rightarrow \infty$) behavior 
of the spatial component of the gauge potential $A_\phi(r)$, respectively $A_y(r)$, being of the
form $A_i(r)=S_i-J_i/r^2$ are related to the superfluid velocity $S_i$ and superfluid current
$J_i$ in the dual field theory.

In agreement with the previous analysis of the similar models without superflow
we have found that the transitions are of the second order, as indicated by the
mean field values of the critical exponents. The presence of the superflow does not
change the fact that, neither the critical value of the chemical potential nor
the density operator  depend on $\alpha$-coupling constant. 
The spatial components of the Maxwell
field identified with the currents at the boundary are affected by the coupling 
to {\it dark matter} sector. Also the current of the s-wave holographic superconductor 
in the solitonic background (equation (\ref{curr-sfi})) does not depends on the {\it dark matter}
sector. Its dependence on the chemical potential is of the form $J_\phi \propto (\mu - \mu_c)$.

The critical chemical potential of the SU(2) p-wave holographic superconductor/superfluid 
in the solitonic background is strongly effected by the presence of {\it dark matter}. Not only 
it depends on the velocity of the condensate $S_\phi$  but also the coupling $\alpha$, 
 the {\it dark matter} chemical potential $\mu_D$ and the {\it dark matter} velocity $S_D$. These dependencies
have been illustrated in figures \ref{fig3} and \ref{fig4}.

The black hole taken as a gravitational 
background allows the study of finite temperature phase transitions. The critical temperature
of the model with superflow has to be calculated self-consistently, due to the transcendental 
character of the corresponding equations. This is true for both s-wave and p-wave symmetries.
The s-wave superconducting transition temperature does not depend on the coupling constant to
the {\it dark matter} sector. Contrary to that, the transition temperature of the p-wave SU(2) holographic
superconductor depends on the {\it dark matter} not only $via$ the coupling $\alpha$ 
but also through the {\it dark matter} velocity $S_D$ and density $\rho_D$. In the figures, the last two 
parameters have been quantified by their ratio to corresponding parameters of the Maxwell sector.

In $SU(2)$ p-wave superfluids one can observe  an interesting duality between solitonic and black hole backgrounds. 
Near $z \rightarrow 0$, in the case of solitonic background 
one has that $A_\phi$ is function of $z^2$ and $A_t$ tends to the constant value 
of chemical potential $\mu$.
On the contrary, in black hole background the situation changes dramatically. 
The $t$-component is function of $z^2$, while the spatial one leads to the constant 
value of the superfluid velocity $S_y$.

The table I summarizes our findings related to the dependence on the {\it dark matter} of the critical
chemical potentials, critical temperatures, the order parameters and the currents
for both symmetries and both gravitational backgrounds, respectively. Of some interest are the universal relations
of the currents on $\mu$, which read $J \propto (\mu-\mu_c)$ for both s-wave and p-wave 
holographic superconductors.  Much more important is the dependence of the currents on temperature.
Again for both symmetries we get the linear in $(1-T/T_c)$ dependence. This important finding 
which might be of relevance  for real superconductors, as discussed below. In the literature, 
the above behavior is sometimes referred as the Onnes relation~\cite{kvit15,dew01}.

\red{
\begin{table}
\caption{\label{tableone} Summary of the obtained results.  The table shows
the functional dependence of the critical chemical potentials and temperatures as
well as order parameters and currents of holographic superconductors on
dark matter parameters $\alpha$, $\mu_D$ and $S_D$ (see text). Note
the linear dependence of the currents on temperature for both symmetries of the
order parameter independently of its symmetry.}
\begin{center}
\begin{tabular}{|c|c|c|}
\hline
 &  soliton (T=0)          & black hole ($T\ne 0$) \\
 \hline
 & $\mu_c(\alpha)=\mu_c(0)$ &               $T_c=T_c(\alpha)$   \\
s-wave & $<\cO(\alpha)>=\sqrt{\tilde{\alpha}}<\cO(0)>\propto (\mu-\mu_c)$ &    $<\cO(\alpha)>=\sqrt{\tilde{\alpha}}<\cO(0)>$    \\
 & $J_\phi(\alpha)=J_\phi(0)(\mu-\mu_c)$ & $J_y(\alpha,T)=J_y(T_c)(1-\frac{T}{T_c})$ \\
\hline
         & $\mu_c=\mu_c(\alpha,\mu_D,S_D)$ & $T_c=T_c(\alpha,\mu_D,S_D)$ \\
 p-wave  & $<\cO(\alpha)>=\sqrt{\tilde{\alpha}}<\cO(\mu_c)>$ &   $<\cR(\alpha)>=\sqrt{\tilde{\alpha}}<\cR(T_c)>$  \\
 SU(2)	 & $J_\phi(\alpha)=J_\phi(\mu_c)(\mu-\mu_c)$ & $J_y(\alpha,T)=J_y(T_c)(1-\frac{T}{T_c})$     \\
\hline
\end{tabular}
\end{center}
\end{table}
}

Now we shall discuss the obtained results in the light of experimental data
on different families of superconductors with various symmetries of the order parameter. 
We start with some preliminaries. It is well known that 
the vanishing of the electrical resistivity and the appearance of the ideal diamagnetism
($i.e.,$ the expulsion of the external magnetic field) below specific transition 
temperature $T_c$ are two defining characteristics of superconductors, with the latter 
being of utmost importance for the understanding of the phenomenon.
The real material remains in the superconducting phase if its  temperature $T$, magnetic field $B$ 
and the current  (density) $J$ are kept below their
critical values, $T_c$, $B_c$ or $J_c$, respectively. It means that
on phase diagram in the 
temperature $T$, magnetic field  $B$ and current $J$ space, there exist critical surface below which
the system is  superconducting.  
Out of the range of the aforementioned parameters, the changes of the superconducting properties of materials, subject to the current flow, are not quite well
theoretically elaborated.
However, it is known
that if the current flowing in a superconductor exceeds a certain critical value, the system
undergoes a superconductor to the normal metal transition.  

The response  of the superconductor  to the current flow is 
completely different for the type I and type II superconductors, due to the appearance and the flow of vortices
 and the concomitant existence of two critical fields (lower and upper), in the latter.
Generally it has been assumed that the superconductors undergo superconductor - normal conductor
transition, if the flowing current produces on the surface of material magnetic field of the
order of the critical one $H_c$~\cite{cyrot-pavuna}. Type II superconductors are typically used
for applications and they also can sustain only the finite currents. The maximum value of $J_c$ is
related to the lower critical field $H_{c1}$ as indicated in recent experiments 
studying the properties of current currying superconductors~\cite{tal15,tal16,kam16}, where
some universalities observed in a number of different families 
of superconductors with various symmetries of the order parameter have been pointed out. 
Among all, it has been shown that for thin films of thickness $b$ less
than the penetration depth $\lambda$, there exist a limiting value of the current $J_c$ which for type I superconductors
is $H_c/\lambda$, whereas for type II materials  $H_{c1}/\lambda$, where $H_{c1}$ is the lower critical field~\cite{tal15}.

More recent data seem to indicate that the relation between $J_c$ and the penetration depth 
changes from $J_c\propto \lambda^{-3}$ valid for films with $b \le \lambda$ to  $J_c\propto \lambda^{-2}$ valid 
for films with $b >>\lambda$. This analysis which the main aim  was to show the above dependence 
in the whole temperature range and  for a number of
different superconductors with various symmetries of the order parameter, obtained an additional 
support from the present ond  earlier calculations using holography.
Namely, the 
holographic analysis of (2+1)-dimensional superconductors and the present one studying 3+1 dimensional systems
show universal temperature dependence of the currents for s-wave and p-wave superconductors 
of the form $J(T)=J(0)(1-T/T_c)^\nu$ with the expenent $\nu$ depending only on the dimensionality of the system
with $\nu=3/2$ for two dimensional and $\nu=1$ for three dimensional systems.
The experimental data on three dimsional sample show~\cite{kvit15} $J(T)=J(0)(1-T/T_c)$ 
temperature dependence for $T$ close to $T_c$. 
For superconducting films of the thickness lower than the penetration depth, the experimental data
for the critical current close to $T_c$ well agree with the dependence given by the relation $J(T)=J(0)(1-T/T_c)^{3/2}$.
Recent summary of a number of experimental data~\cite{tal16} supports the holographic results.

\acknowledgments
MR was partially supported by the grant no. $DEC-2014/15/B/ST2/00089$ of the National Science Center 
 and KIW by the grant DEC-2014/13/B/ST3/04451.


\end{document}